\documentclass[aps,twocolumn,superscriptaddress,groupedaddress,10pt,nofootinbib]{revtex4} 

\usepackage{graphicx}  	
\usepackage{dcolumn}   	
\usepackage{bm}        	
\usepackage{amssymb}   	
\usepackage{amsmath}
\usepackage{epstopdf}
\usepackage[a4paper,margin=20mm]{geometry}
\usepackage{color}  

\begin{document}

\title{Low-Reynolds-number, bi-flagellated Quincke swimmers with multiple forms of motion}

\author{Endao Han}
\affiliation{Joseph Henry Laboratories of Physics, Princeton University}
\author{Lailai Zhu}
\affiliation{Department of Mechanical Engineering, National University of Singapore}
\author{Joshua W. Shaevitz} 
\affiliation{Joseph Henry Laboratories of Physics, Princeton University}
\affiliation{Lewis-Sigler Institute for Integrative Genomics, Princeton University}
\author{Howard A. Stone} 
\email[E-mail:]{hastone@princeton.edu}
\affiliation{Department of Mechanical and Aerospace Engineering, Princeton University}

\date{\today}

\begin{abstract}

In the limit of zero Reynolds number (Re), swimmers propel themselves exploiting a series of non-reciprocal body motions. 
For an artificial swimmer, a proper selection of the power source is required to drive its motion, in cooperation with its geometric and mechanical properties. 
Although various external fields (magnetic, acoustic, optical, etc.) have been introduced, electric fields are rarely utilized to actuate such swimmers experimentally in unbounded space. 
Here we use uniform and static electric fields to demonstrate locomotion of a bi-flagellated sphere at low Re via Quincke rotation. 
These Quincke swimmers exhibit three different forms of motion, including a self-oscillatory state due to elasto-electro-hydrodynamic interactions. 
Each form of motion follows a distinct trajectory in space. 
Our experiments and numerical results demonstrate a new method to generate, and potentially control, the locomotion of artificial flagellated swimmers. 

\end{abstract}

\maketitle


In a Newtonian fluid, locomotion of micro-swimmers requires non-reciprocal body motions \cite{Taylor_1951, Purcell_1977, Lauga_review}. Bacteria or eukaryotic cells achieve this by beating or rotating their slender hair-like organelles, flagella \cite{Berg_flagella, Berg_review} or cilia \cite{Goldstein_ARFM}, powered by molecular motors. 
Mimicking these organisms, artificial swimmers propelled by rotating helices \cite{Helical_2009, Nelson_2009} or whipping filaments \cite{Stone_2005, Goldstein_1998, Hosoi_2006, HybridSwimmer} have been fabricated. They are commonly driven by an external power source such as a magnetic field \cite{Stone_2005, Helical_2009,Nelson_2009,Box_2017, AdaptiveSwimmer}, sound \cite{Acoustic_2016}, light \cite{Light_2009,Light_2016}, biological materials \cite{HybridSwimmer}, etc. 
However, there are very few electrically powered micro-swimmers \cite{Loget_2011}, although electric fields have been exploited to drive other active systems \cite{Lobry_1999,Lemaire_2007_PRL,Bartolo_2013, Brosseau_2017,karani2019tuning,Lauga_2019} via a phenomenon called Quincke rotation \cite{Quincke_1896}.

\begin{figure}[b!]
    \begin{center}
        \includegraphics[scale = 0.95]{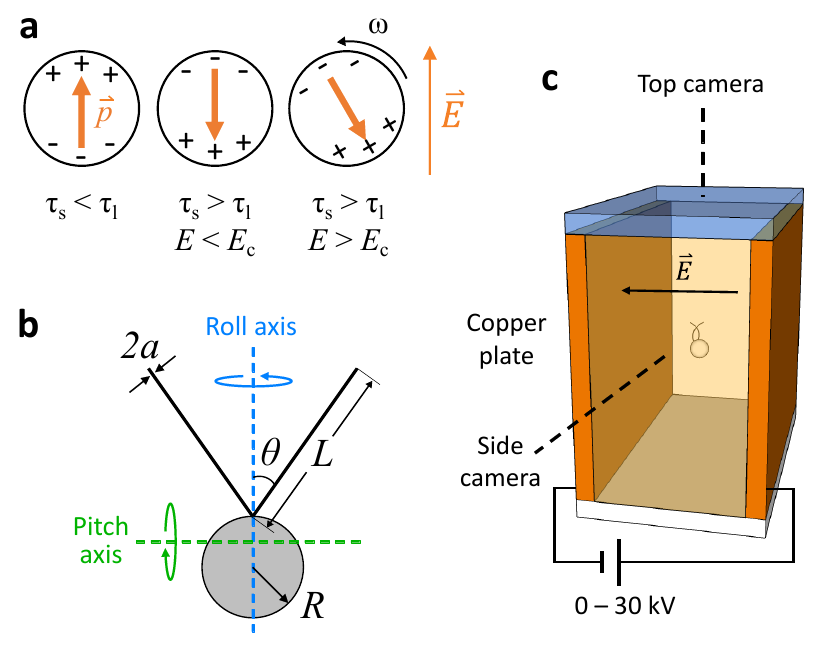}
    \end{center}
\caption{ Quincke rotation and the experimental setup. 
(a) Distribution of free charge and the corresponding dipole $\vec{p}$ on a sphere in a uniform, steady electric field $\vec{E}$. 
From left to right, the sphere is stationary, stationary, and rotating with a constant angular velocity $\omega$, respectively. 
(b) A sketch of the bi-flagellated swimmer. Dashed lines show the roll axis (blue) and pitch axis (green). 
(c) A schematic illustration of the experimental setup. }
\label{ExptSetup}
\end{figure}

Quincke rotation originates from an electro-hydrodynamic instability \cite{Tsebers_1980,Jones_1984,Cebers_2001}. 
Submerged in a liquid with permittivity $\varepsilon_\text{l}$ and conductivity $\sigma_\text{l}$, a spherical particle with permittivity $\varepsilon_\text{s}$ and electric conductivity $\sigma_\text{s}$ is polarised under a uniform, steady electric field $\vec{E}$. 
When the particle is stationary, the induced dipole $\vec{p}$ due to the free charges is parallel or antiparallel to $\vec{E}$ (Fig.~\ref{ExptSetup}(a)): 
If the particle's relaxation time $\tau_\text{s} = \varepsilon_\text{s} / \sigma_\text{s}$ is shorter than that of the ambient liquid, $\tau_\text{l} = \varepsilon_\text{l} / \sigma_\text{l}$, $\vec{p}$ points in the same direction as $\vec{E}$; when $\tau_\text{s} > \tau_\text{l}$, $\vec{p}$ is opposite to $\vec{E}$, which generates an electric torque $\vec{\Gamma}_\text{Q} = \vec{p} \times \vec{E}$ that amplifies any angular perturbation. 
However, due to the resisting viscous torque $\vec{\Gamma}_\mu$, the system becomes unstable only when $E = |\vec{E}|$ exceeds a threshold $E_\text{c}$. 
This instability causes the particle to rotate with a constant angular velocity $\omega$:  
\begin{equation}
    \omega = \frac{1}{\tau} \sqrt{\left( \frac{E}{E_\text{c}} \right)^2-1}, 
    \label{eq:AngularSpeed}
\end{equation}
where $\tau = \dfrac{\varepsilon_\text{s}+2\varepsilon_\text{l}}{\sigma_\text{s}+2\sigma_\text{l}}$ is the relaxation time of the system (see Supplemental Information for derivation), and the rotational axis can be in any direction perpendicular to $\vec{E}$. 
During steady-state Quincke rotation, there is a constant angle between $\vec{p}$ and $\vec{E}$ (Fig.~\ref{ExptSetup}(a)), which results in a non-zero $\vec{\Gamma}_\text{Q}$.

\begin{figure*}[ht!]
    \begin{center}
        \includegraphics[scale = 0.95]{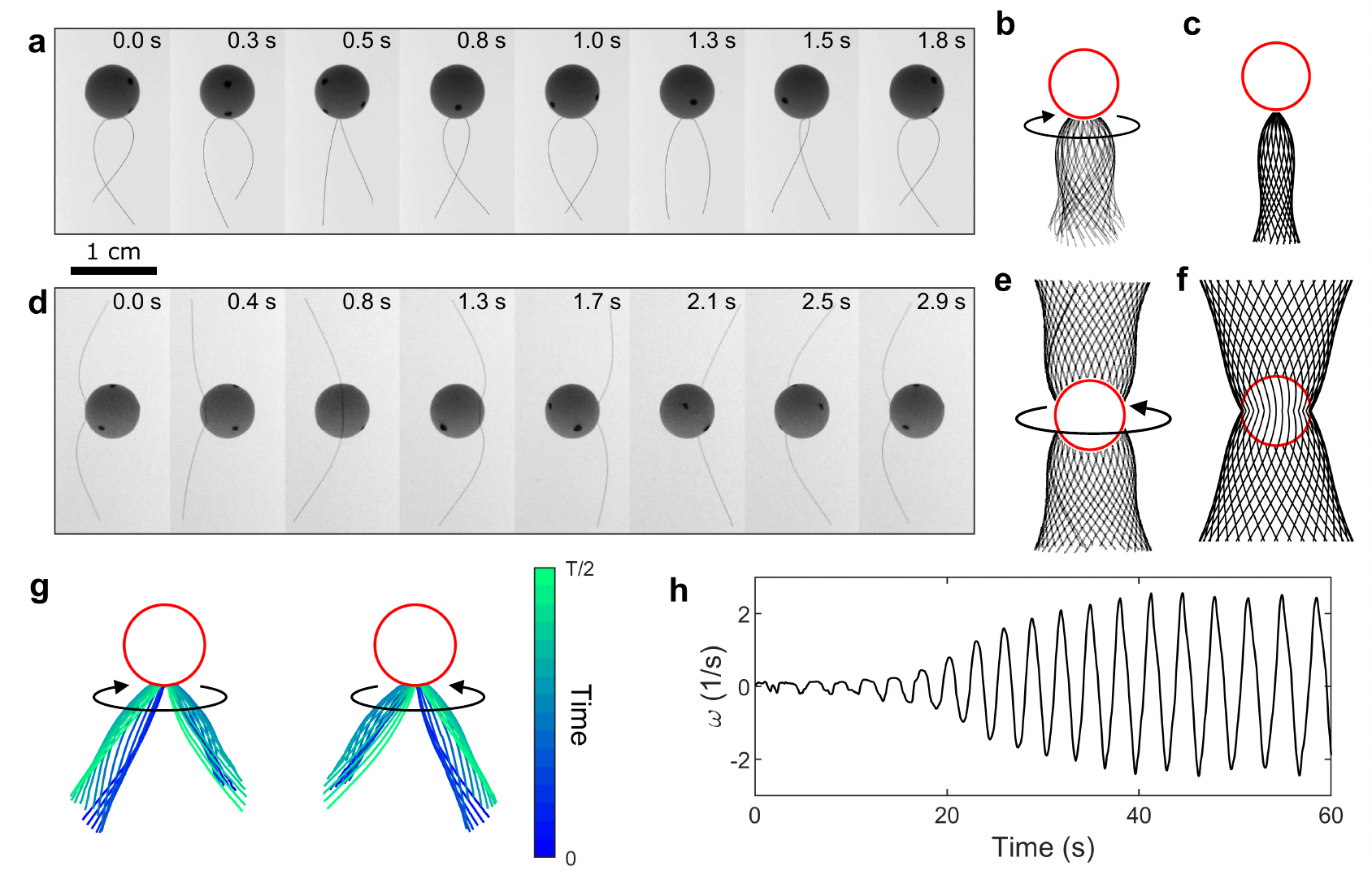}
    \end{center}
\caption{ Bi-flagellated swimmers driven by a uniform, steady electric field in the horizontal direction. 
(a) Rotation (in one period $T$) of a swimmer with $\theta = 47^\circ$ around its roll axis in the co-moving frame using the center of the sphere as the reference, and the overlapped tail shapes obtained from the experiments (b) and numerical simulations (c). 
The scale bar indicating 1~cm is for both (a) and (d). 
(d) Rotation (in $T$) of a swimmer with $\theta = 58^\circ$ around its pitch axis in the co-moving frame, and the corresponding experimental (e) and numerical (f) tail shapes. 
The tail profiles inside the projection of the sphere are omitted in (e) due to visualization difficulties. 
(g) Experimental time lapse shapes in the co-moving frame of the swimmer ($\theta = 44^\circ$) undergoing a self-oscillatory motion . 
The left and right images correspond to a clockwise rotation (viewing from the top) at time $t \in [0, T/2]$ and a counter-clockwise rotation at $t \in [T/2$ to $T]$, respectively.
(h) Instantaneous angular velocity $\omega$ as a function of time for the swimmer with $\theta = 44^\circ$ at $E = 1.65 E_\text{c}$ showing how it reached a steady oscillatory state from being stationary initially. 
}
\label{SwimmerImages}
\end{figure*}

Recently, a flagellated swimmer in unbounded space driven by Quincke rotation has been proposed theoretically \cite{Lailai_2019,Lailai_2020}. 
In light of the theory, we built a laboratory prototype---a bi-flagellated Quincke swimmer comprised of a spherical particle and two attached elastic filaments, as shown in Fig.~\ref{ExptSetup}(b), and systematically studied its behaviors at low Reynolds number ($\mathrm{Re}<0.3$, see Methods). 
Varying the electric field $\vec{E}$ and the angle between the two filaments, the Quincke swimmers exhibit three distinct forms of motion---two unidirectional rotations, which we call roll and pitch, and a self-oscillatory rotation, due to the balances between the electrical, elastic, and hydrodynamic torques, resulting in distinct trajectories in space. 
Surprisingly, a recent work \cite{Bente_2020_eLife} reported that spherical bacteria \textit{Magnetococcus marinus} exhibit a similar pitch motion as our bi-flagellated artificial swimmers, which is rarely adopted by other microorganisms. 
Moreover, we found a threshold tail angle that separates the swimmers' preferred forms of rotation, and within a small range close to this threshold angle, the three forms of motion coexist. 


\subsection*{Experimental setup}
Each swimmer was comprised of a spherical particle and two symmetric tails (Fig.~\ref{ExptSetup}(b)). 
We tested swimmers with two different sizes: 
The small swimmers had a particle radius $R = 1.59~$mm and a tail radius $a = 13.7~\mu$m, while the big swimmers had $R = 3.18~$mm and $a = 24.8~\mu$m. 
In this paper, we report results obtained with the big swimmers that had a fixed tail length $L = 13.6 \pm 0.6~$mm ($L/R = 4.3 \pm 0.2$ and $a/L = (1.8 \pm 0.1) \times 10^{-3}$), and focus on one geometric control parameter --- the angle $\theta$ between the tails and the symmetry axis. 
The experimental setup is illustrated schematically in Fig.~\ref{ExptSetup}(c). 
A constant and uniform electric field was generated by a pair of parallel copper plates attached to the inner walls of an acrylic container filled with a weakly conductive oil. 
The density of the oil approximately matched that of the spheres, and its viscosity was $\mu = 0.225$~Pa$\cdot$s. 
The intensity of the electric field ranged from 0 to $7.7 \times 10^5$~V/m. 
We captured and reconstructed the three-dimensional motion of the swimmers with two cameras imaging from perpendicular directions. 
Details on the experimental setup, material properties, and image analysis techniques are described in the Methods.

\subsection*{Multiple forms of motion}

Given a sufficiently large $E$, the bi-flagellated swimmers displayed three forms of motion as shown in Fig.~\ref{SwimmerImages}. 
As the sphere rotated, the tails were deformed by the hydrodynamic forces.
Their profiles depended on the geometry of the swimmer: 
When $\theta$ was below a threshold $\theta_\text{c}$, the swimmer rotated about its axis of symmetry (roll axis in Fig.~\ref{ExptSetup}(b)) with the tails buckled inward, forming a helical shape (Fig.~\ref{SwimmerImages}(a)), and generating thrust akin to the observations of Refs.~\cite{Manghi_PRL,Breuer_2008,Fermigier_2008,Reis_2015}. 
A swimmer with $\theta > \theta_\text{c}$ rotated about an axis perpendicular to its axis of symmetry (pitch axis in Fig.~\ref{ExptSetup}(b)), and its tails buckled outward (Fig.~\ref{SwimmerImages}(d)). 
The corresponding overlaid images of the rotating tails within one period are shown in Fig.~\ref{SwimmerImages}(b) and (e), respectively. 
Furthermore, we conducted simulations based on \cite{Lailai_2019,Lailai_2020} that reproduced these two motions, as presented in Fig.~\ref{SwimmerImages}(c) and (f). 

Within a small range of $\theta$ and $E$, the swimmer underwent an oscillatory rotation  (Fig.~\ref{SwimmerImages}(g)) in contrast to the classical, unidirectional Quincke rotation. 
Fig.~\ref{SwimmerImages}(h) shows the particle's angular velocity $\omega$ as a function of time when the amplitude of the oscillation increased away from zero toward a plateau. 
The occurrence of such a cyclic behavior driven by a steady electric field indicates its self-oscillatory nature as opposed to a forced oscillation. 
The self-oscillatory rotation emerges from an elasto-electro-hydrodynamic instability through a Hopf bifurcation, as identified theoretically for a similar system~\cite{Lailai_2019,Lailai_2020}. 
Transient oscillations also appeared in the experiments during the early stages of continuous rotations, but as the amplitude increased, eventually, the restoring force provided by the tails was not sufficient to sustain the oscillation, and $\omega$ became a constant. 
In the simulations, we observed not only this oscillatory state about the roll axis but also an oscillation about the pitch axis, when the swimmer's axis of symmetry was initially parallel to $\vec{E}$. 
However, the latter was absent in our experiments.

\subsection*{Discontinuous transition in angular speed and hysteresis}

\begin{figure*}[ht!]
    \begin{center}
        \includegraphics[scale=0.95]{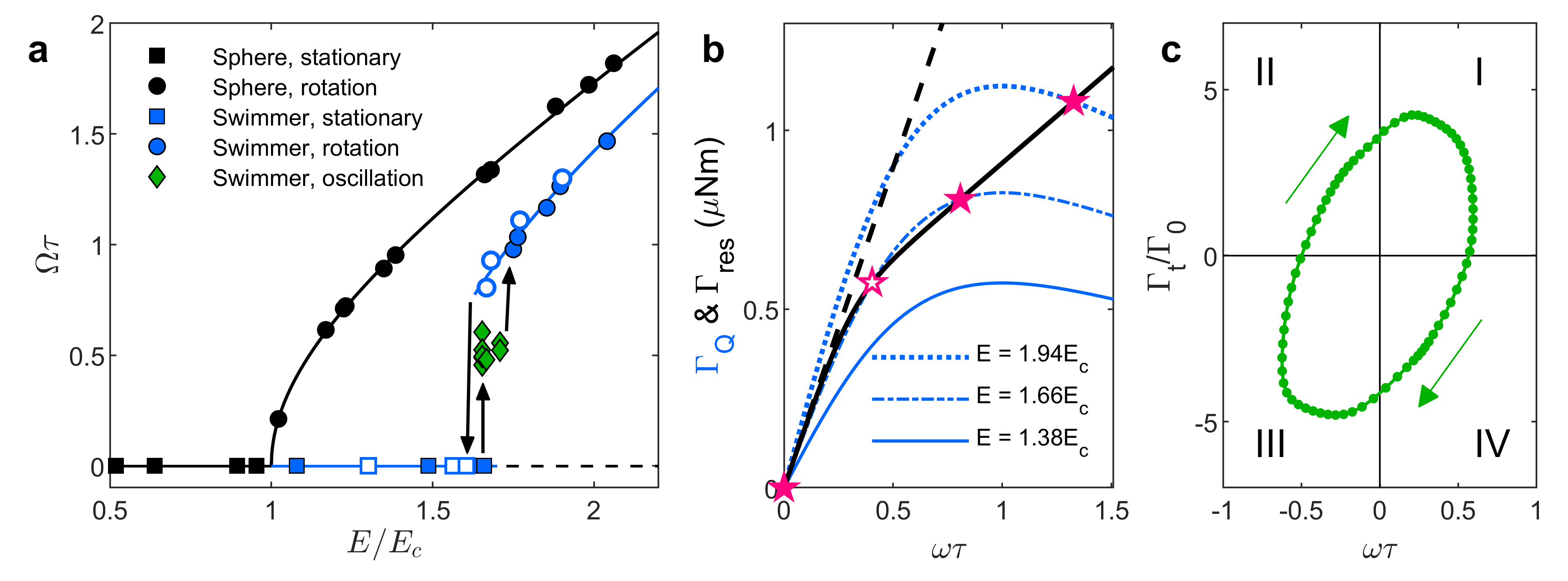}
    \end{center}
\caption{  Experimentally measured angular speed of the swimmer and the corresponding torques. 
(a) Nondimensional angular frequency $\Omega \tau$ of a particle and a swimmer with $\theta = 47^\circ$ when the field intensity $E$ increased (solid points) or decreased (open points), with data points labelled in the panel. The solid black curve shows Eq.~(\ref{eq:AngularSpeed}). 
The solid blue curve is a guide to the eye. 
The black arrows show how the state of the swimmer changes as $E$ increases or decreases. 
(b) The driving torque $\Gamma_\text{Q}$ (Eq.~\ref{eq:T_Q}) at three different $E$ (blue lines) and the resisting torque $\Gamma_\text{res} = |\Gamma_\mu+\Gamma_\text{ec}+2\Gamma_\text{t}|$ (solid black line, see Supplemental Information for details) of the swimmer while rolling. The dashed black line shows the slope of $\Gamma_\text{res}(\omega)$ as $\omega \to 0$. The stable fixed points are labelled with solid stars and the unstable fixed points with the open star. 
(c) The trajectory of torque-angular velocity when the swimmer oscillated. The trajectory is calculated using Eqs.~\ref{eq:Rotation} and \ref{eq:T_Q} (see Supplemental Information for details). Following \cite{Fermigier_2008}, we scale $\Gamma_\text{t}$ by $\Gamma_0 \equiv \kappa/L$, where $\kappa$ is the bending stiffness of the tails. The green arrows label the direction of time. }
\label{AngularSpeed}
\end{figure*}

To understand the dynamics of the system, we look at the swimmer with $\theta = 47^\circ$, which is slightly below $\theta_\text{c}$. 
We measured the average angular frequency of the particle $\Omega \equiv 2\pi/T$, with $T$ being one period of oscillation or rotation. 
For unidirectional rotation (roll or pitch), the instantaneous angular speed is $|\omega| = \Omega$, but for oscillation, $\omega \approx \omega_0 \text{sin}(\Omega t)$, where $\omega_0$ is the amplitude of $\omega$. 
We plot the dimensionless angular frequency $\Omega\tau$ measured using the relaxation time $\tau$ versus $E$ in Fig.~\ref{AngularSpeed}(a) for both a bare sphere and the swimmer.
The rotation of the sphere is well described by Eq.~(\ref{eq:AngularSpeed}).
In comparison, as we increased $E$, the swimmer was stationary until $E$ reached about $1.7 E_\text{c}$, where the oscillatory state occurred within a narrow window. 
When a stronger electric field was applied, the rotation about the roll axis became unidirectional and steady, and the angular frequency $\Omega$ increased with $E$.
When we reversed this process by decreasing $E$, hysteresis emerged and no stable oscillatory state was observed. 

To explain the discontinuous transition and hysteresis in Fig.~\ref{AngularSpeed}(a), we examine the torques applied on the particle. 
Besides the driving torque generated by the Quincke effect $\Gamma_\text{Q}$ and the viscous torque due to the spinning sphere $\Gamma_\mu = -8 \pi \mu R^3 \omega$, each tail exerted a same torque $\Gamma_\text{t}$ on the particle since they were symmetric. 
When the swimmer pitched, the drag force generated a torque $\Gamma_\text{ec}$ due to the sphere's eccentric rotation, which was negligible for rolling and oscillation. 
The total torque vanishes in the low Re limit, thus 
\begin{equation}
    \Gamma_\text{Q} + \Gamma_\mu + \Gamma_\text{ec} + 2\Gamma_\text{t} = 0.  
    \label{eq:Rotation}
\end{equation}
Given $E$ and $\omega(t)$, we calculated $\Gamma_\text{Q}$ and used Eq.~(\ref{eq:Rotation}) to solve for $\Gamma_\text{t}$. 


When the swimmer is stationary or rotates steadily, the driving torque is 
\begin{equation}
    \Gamma_\text{Q} = 8\pi \mu R^3 \left( \frac{E}{E_c} \right)^2 \frac{\omega}{1+(\omega \tau)^2}, 
    \label{eq:T_Q}
\end{equation}
shown by the blue curves in Fig.~\ref{AngularSpeed}(b) (see derivation in Supplemental Information). 
If the tails of the swimmer are rigid, the resisting torque $\Gamma_\text{res} = |\Gamma_\mu+\Gamma_\text{ec}+2\Gamma_\text{t}|$ is a linear function of $\omega$ following the dashed black line in Fig.~\ref{AngularSpeed}(b). 
The system evolves toward where $\Gamma_\text{Q}$ and $\Gamma_\text{res}$ intersects (fixed points). 
Consequently, as $E$ increases, the transition occurs via a supercritical pitchfork bifurcation, as in the original Quincke rotation. 
However, since the elastic tails deform under large torques (Fig.~\ref{SwimmerImages}), $\Gamma_\text{res}$ deviates from a straight line and bends down as $\omega$ increases (see Supplemental Information), leading to a subcritical bifurcation. 
At small $E$, the point $(0,0)$ in Fig.~\ref{AngularSpeed}(b) is a stable fixed point. 
The swimmer remains stationary until $\text{d}\Gamma_\text{Q}/\text{d}\omega$ exceeds $\text{d}\Gamma_\text{res} / \text{d}\omega$ at $\omega = 0$ as $E$ increases, where the origin becomes unstable and the stable fixed point shifts discontinuously to $\omega \ne 0$.  
For an intermediate $E$, e.g., $E = 1.66 E_\text{c}$, there are two stable fixed points (one of them is at $(0,0)$) separated by an unstable fixed point. 
As a result, the swimmer stayed at different states when $E$ increased or decreased, hence the hysteresis. 

When the swimmer oscillates (Fig.~\ref{SwimmerImages}(g)), both $\Gamma_\text{Q}$ and $\Gamma_\text{t}$ do not solely depend on the instantaneous $\omega$, but also on its preceding values.
The trajectory of $\Gamma_\text{t}(\omega)$ in Fig.~\ref{AngularSpeed}(c) tends to a limit cycle instead of a fixed point (see Supplemental Information). 
$\Gamma_\text{t}$ accelerated the rotation in the I and III quadrants and decelerated it in the II and IV quadrants of the figure. 
In each period of oscillation, $\Gamma_\text{t}$ was strong enough to hinder the rotation, and then the residual elastic energy stored in the tails drove the sphere to rotate in the opposite direction. 

\subsection*{State diagram}

\begin{figure}
    \begin{center}
        \includegraphics[scale=0.95]{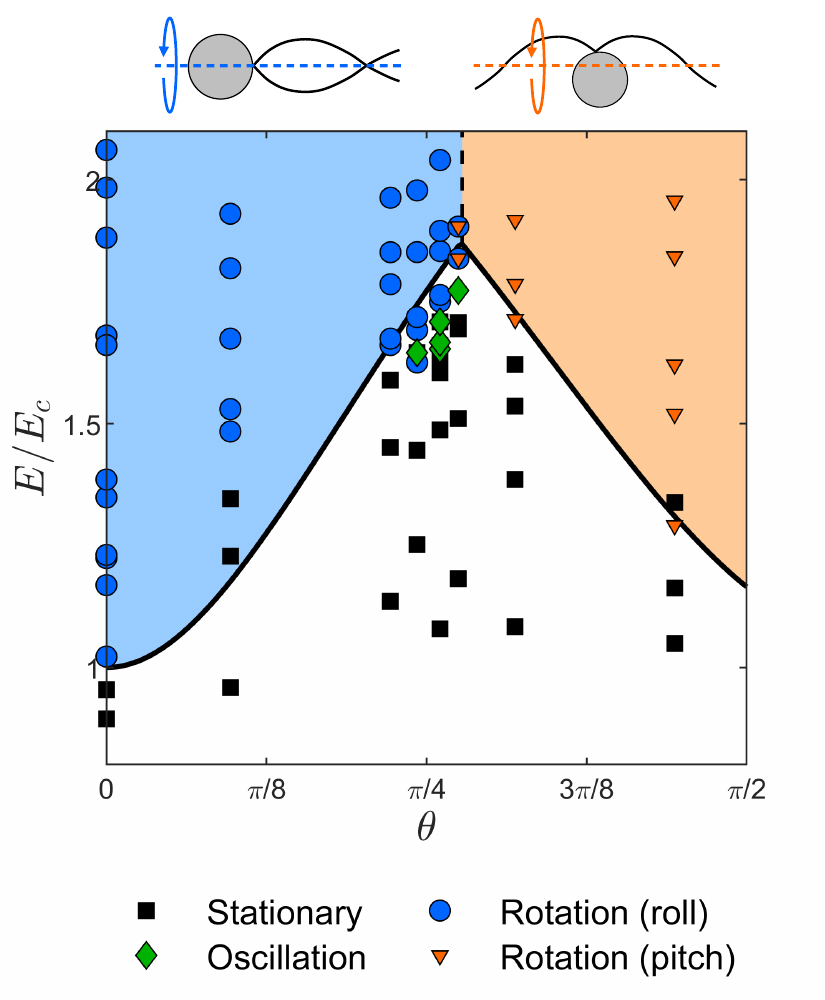}
    \end{center}
\caption{ State diagram for swimmers with $L/R = 4.3$ and different $\theta$. The solid black curve shows the calculated onset field at different $\theta$ without any fitting parameters. The dashed black line indicates the boundary of the roll (left) and pitch (right) states at $\theta_\text{c} = 49.8^\circ$. 
}
\label{StateDiagram}
\end{figure}

The form of motion of a swimmer, or its ``state'' (stationary, rolling, pitching, or oscillatory), was mainly controlled by its geometry and the applied field. 
Fixing the ratio of tail length to particle size $L/R = 4.3$, we mapped out the bi-flagellated Quincke swimmers' state diagram with respect to the tail angle $\theta$ and the relative applied field $E/E_\text{c}$, as shown in Fig.~\ref{StateDiagram}. 
We predict the swimmers' motion by identifying the axis around which it experiences the least viscous torque, assuming the tails are both rigid. 
We then numerically calculated the threshold electric field $E_\theta$ for continuous rotation for different $\theta$, shown by the black curve in Fig.~\ref{StateDiagram}, with no adjustable parameter (see Supplemental Information). 
The calculations agree reasonably well with the experimental observations\footnote{One reason why the predicted onset field deviates from experimental measurements (e.g. at $\theta = 0.3 \pi$) could be that we assumed the bases of the two tails were next to each other in the calculation to keep the model simple, while in actual experiments there was, on average, an approximately 1.8~mm gap between them.}. 

The boundary separating the rolling and pitching motions is at $\theta_\text{c} = 49.8^\circ$. 
Swimmers with $\theta < \theta_\text{c}$ preferred rolling, and those with $\theta > \theta_\text{c}$ preferred pitching. 
Within a small region where $\theta \in (43^\circ,~ \theta_\text{c})$ and $E$ was slightly below $E_\theta$, the swimmers exhibited stable oscillatory rotations. 
When $\theta \approx \theta_\text{c}$, the swimmer was almost equally likely to rotate about any axis (see Supplemental Information), so the three forms of motion, roll, pitch, and oscillation, coexisted. 
In this case, besides the field intensity $E$, the eventual stable form of motion was significantly affected by the initial orientation of the swimmer relative to $\vec{E}$. 
As introduced above, Quincke rotation can occur around any axis perpendicular to the external field. 
Consequently, if the symmetry axis of the swimmer was perpendicular to $\vec{E}$, it tended to roll, while if the axis was parallel to $\vec{E}$, it tended to pitch.

\subsection*{Translational motion and propulsive force}

\begin{figure*}
    \begin{center}
        \includegraphics[scale = 0.9]{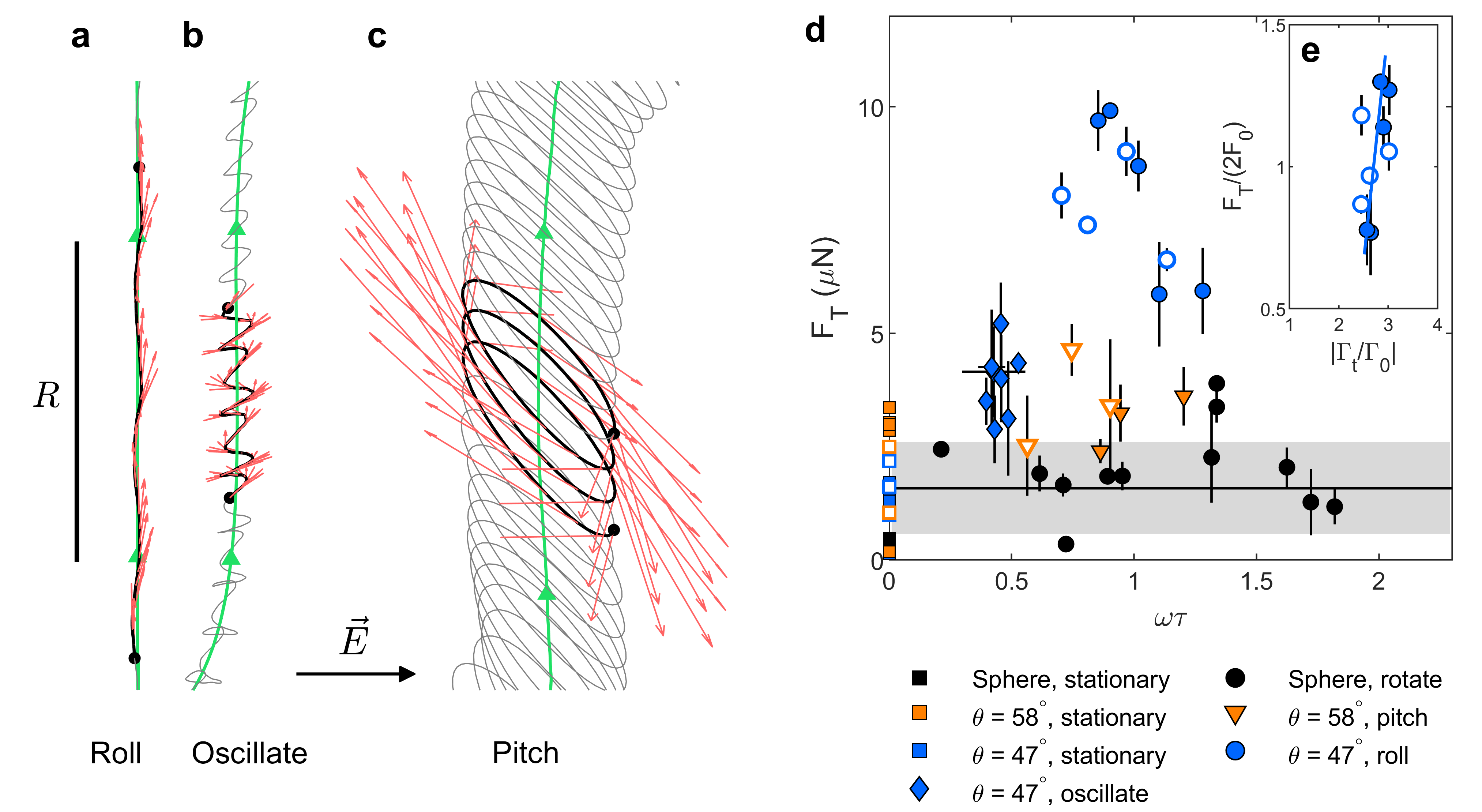}
    \end{center}
\caption{ Trajectories and propulsive forces generated by different forms of motion. 
(a-c) Trajectories of the particle and the propulsive force vectors at each time point (red arrows). 
The forms of motion were (a) roll ($\theta = 40^\circ$), (b) oscillation ($\theta = 44^\circ$), and (c) pitch ($\theta = 58^\circ$). 
The thick black curves highlight three periods of rotation or oscillation in each plot, with the two ends indicated by black dots. 
The green curves show the smoothed trajectories, and the directions of motion are labelled by the green arrow heads. 
The views are perpendicular to the trajectories in every plot, and the scale bar indicating the particle radius $R$ is shared in (a-c). 
The direction of the external field $\vec{E}$ for all three panels is labelled by the black arrow between (b) and (c). 
(d) The propulsive force averaged over each period $F_\text{T} = \left< \frac{1}{T} \left| \int_t^{t+T} \vec{F}_\text{2t}(t) \mathrm{d}t \right| \right>$ as a function of the nondimensional angular speed. 
Solid symbols represent increasing $E$ and open symbols represent decreasing $E$. 
(e) Relationship between the propulsive force generated and torque applied on each tail. The forces and torques are nondimensionalized with $F_0 = \kappa/L^2$ and $\Gamma_0 = \kappa/L$, respectively, where $\kappa$ is the bending stiffness of the tail \cite{Fermigier_2008}. 
}
\label{Trajectory_Force}
\end{figure*}

Lastly, we studied the locomotion of the swimmers while executing different forms of motion. 
For each swimmer, the translational motion of the particle is determined by three forces: the viscous drag $\vec{F}_\mu = -6 \pi \mu R \vec{v}$, the gravitational force due to a slight density mismatch $\vec{F}_\rho = \frac{4}{3} \pi R^3 (\rho_\text{s} - \rho_\text{l}) \vec{g}$, and the propulsive force $\vec{F}_\text{2t}$ provided by both tails. 
Here $\vec{v}$ is the velocity of the particle, $\vec{g}$ is the gravitational acceleration, and $\rho_\text{s}$ and $\rho_\text{l}$ are the densities of the particle and the liquid, respectively.  
These three forces add up to zero, so $\vec{F}_\text{2t} = -\vec{F}_\mu - \vec{F}_\rho$. 
Typical trajectories corresponding to rolling, oscillation, and pitching are presented in Fig.~\ref{Trajectory_Force}(a-c), respectively, along with the instantaneous propulsive force $F_\text{2t}$. 
While rolling, the swimmer moved along a smooth and relatively straight trajectory (Fig.~\ref{Trajectory_Force}(a)), and the force almost always pointed in the same direction as $\vec{v}$. 
When oscillating (Fig.~\ref{Trajectory_Force}(b)), the trajectory became sinusoidal and the direction and magnitude of the propulsive force varied with the translational and angular velocities. 
When pitching, since the rotation was eccentric, the particle followed a helical trajectory resulting from a combination of a fast circular motion and a slow drift (Fig.~\ref{Trajectory_Force}(c)).

To compare the swimmers' ability of generating propulsion under different conditions, in Fig.~\ref{Trajectory_Force}(d) we plot the time-averaged propulsive force generated by the tails in each rotation period $F_\text{T} = \left< \frac{1}{T} \left| \int_t^{t+T} \vec{F}_\text{2t}(t) \mathrm{d}t \right| \right>$. 
Though all three forms of motion were non-reciprocal, only rolling was able to achieve an effective unidirectional translation for the tested swimmers. 
The propulsion peaked when $E$ slightly exceeded $E_\theta$, then $F_\text{T}$ decreased as $\omega$ increased. 
Pitching resulted in poor locomotion because of its symmetry: though the instantaneous force was several times larger compared to the other two forms of motion, they added up to a small value in one cycle of rotation. 
In comparison, oscillation was able to generate a net force along its drift speed, but it was not as efficient, because the component of the force parallel to the drift direction was relatively small, which can be seen in Fig.~\ref{Trajectory_Force}(b). 

Why does $F_\text{T}$ decrease with $\omega$ when the swimmer was rolling? 
We found that $F_\text{T}$ was approximately proportional to the torque applied on each tail $|\Gamma_\text{t}|$ (Fig.~\ref{Trajectory_Force}(e)). 
However, when a stronger electric field was applied, $|\Gamma_\text{t}|$ decreased as the rolling speed $\omega$ increased. 
In general, for a tilted elastic fiber rotating about one fixed end, $\omega(|\Gamma_\text{t}|)$ is an s-shaped function, and the trend of $|\Gamma_\text{t}|$ versus $\omega$ is controlled by a dimensionless bending parameter $B$ (sometimes also known as the sperm number)
\begin{equation}
    B = \frac{4\pi \mu L^4 \omega}{\kappa [\mathrm{ln}(L/a)+1/2]}, 
    \label{eq:Sp}
\end{equation}
where $\kappa$ is the bending stiffness of the fiber \cite{Manghi_PRL, Breuer_2008, Fermigier_2008}. 
We tested swimmers with different length scales (sphere radius, tail radius, tail length, and tail angle) in liquids with different $\tau$, and indeed we found a transition at $B \approx 10^2$: $|\Gamma_\text{t}|$ decreased with $\omega$ when $B < 10^2$ and increased when $B > 10^2$ (see Supplemental Information). 
The swimmer shown in Fig.~\ref{Trajectory_Force} was in the former regime. 

Relating the locomotion with the state diagram (Fig.~\ref{StateDiagram}), we can see that drastic changes in motion can be achieved by adjusting $E/E_\text{c}$ in a small range, especially around $\theta_\text{c}$. 
For example, with $\theta = 49.4^\circ$, the swimmer was stationary at $E = 1.71 E_\text{c}$, oscillated at $E = 1.77 E_\text{c}$, and rolled at $E = 1.84 E_\text{c}$ with a consistent propulsion. 
Moreover, since rolling and pitching coexisted for this swimmer, how it rotated depended on its initial orientation relative to $\vec{E}$ at the moment when $E$ exceeded the threshold. 
All these features could potentially lead to further questions on controlling the motion of such swimmers.

\subsection*{Conclusions}

We created an artificial swimmer driven by constant and uniform external electric fields, exploiting Quincke rotation. 
The swimmer had a rigid spherical body and two elastic filamentous tails, which allowed it to move at low Reynolds number with three different forms of motion: rolling, pitching, and oscillation, controlled by its geometry and the external field. 
Among them, rolling allowed the swimmer to generate steady translational locomotion. 
We discovered a tail angle at which the three forms of motion coexisted. 
Because of the simple structure and driving method of the swimmer, there is a potential to scale up their numbers and scale down their dimensions so that they become a model system for studying collective motion, swarming \cite{Kearns_2010}, or other phenomena in active matter. 

\begin{acknowledgments}
We thank Janine Nunes and Nan Xue for the help with the experiments. 
We thank Ellie Acosta, Benjamin Bratton, Yong Dou, Matthias Koch, and Talmo Pereira for useful discussions. 
E.H. thanks the support by the NSF through the Center for the Physics of Biological Function (PHY-1734030). 
L.Z. thanks the start-up grant provided by the National University of Singapore (R-265-000-696-133). 
H.A.S. thanks the support by the NSF through the Princeton University Materials Research Science and Engineering Center (DMR-2011750). 
The computational work for this article was performed on resources of the National Supercomputing Centre, Singapore (https://www.nscc.sg).
\end{acknowledgments}

\bibliographystyle{unsrt}
\bibliography{QuinckeSwimmer}

\section*{Methods}

\subsection*{Experimental setup and sample preparation}
The experiments were performed with a rectangular acrylic container. 
A constant and uniform electric field was generated by a pair of parallel copper plates attached to the inner walls of the container. 
Each copper plate was 6.3~mm thick, with machined and polished surfaces. 
The top of the container was covered by an acrylic plate to avoid surface flows at the liquid-air interface induced by high electric fields. 
The distance between the surfaces of the two plates was 3.9~cm. 
The other two dimensions of the container were 15~cm (length) and 10~cm (depth). 
This enclosed volume was fully filled with oil. 
The voltage applied on the copper plates was provided by a DC power supply (Micronta) and a DC voltage amplifier (DCH 3034N1, HVPSI). 
The amplifier transformed a 0 to 12~V input voltage into a 0 to 30~kV output voltage. 
Two single-lens reflex (SLR) cameras (Nikon D3300 or Nikon D5100) were used to take videos of the swimmer from two perpendicular directions. 
One camera was placed above the container and the other one on the side, which allowed us to track the motion of the swimmer in three dimensions. 
In each experiment, the voltage applied on the plates was adjusted to an expected value, then we started recording with both cameras simultaneously. 

The spherical heads of the swimmers were made of high density polyethylene (HDPE), with a radius $R = 3.18$~mm and a density $\rho_\text{s} = 0.94 \pm 0.01$~g/cm$^3$. 
The tails were No. 8-0 surgical sutures (from S\&T, nylon, 24.8~$\mu$m in radius). 
We measured the Young's modulus of the fiber $Y = 2.7 \pm 0.5$~GPa. 
When preparing the swimmer, the nylon fiber was cut to a certain length, and its middle point was attached to the HDPE sphere with a small amount of glue (Loctite 401). 
After the glue cured, the fiber was folded symmetrically and formed an angle. 
Colored spots were painted on the surface of the sphere as tracers. 
Their areas were sufficiently small so that they did not affect the electric or hydrodynamic properties of the sphere surface. 

The liquid was an equal mixture by volume of olive oil (Filippo Berio and Spectrum (NF grade)) and castor oil (Alfa Aesar). 
The viscosity of the mixed oil was 0.225~Pa$\cdot$s measured with a rheometer (Anton Paar MCR 301), and its density was $\rho_\text{l} = 0.94$~g/cm$^3$, which approximately matched the density of the HDPE sphere $\rho_\text{s}$. 
The electric properties of the mixed liquid were characterized by measuring the angular speed $\omega$ of HDPE spheres (with no tail attached) under different applied fields $E$. 
Fitting the data with Eq.~(\ref{eq:AngularSpeed}), we obtained $E_c$ and relaxation time $\tau$. 
In the experiments shown here, the mean threshold electric field was $E_c = 382 \pm 8$~kV/m, and the mean relaxation time was $\tau = 0.28 \pm 0.02$~s. 
To prevent the oil from being contaminated by the dust in the air, the whole setup was placed in a glove box, and the oil was filtered on a daily basis. 

Lastly, we estimate the Reynolds number. 
The maximum angular speed of the swimmer did not exceed $\omega \tau = 2$ in our experiments, so we can calculate the upper limit of $\mathrm{Re} = \rho_\text{l} R^2 \omega / \mu$ using $\omega_\text{max} = 7.1$~rad/s. 
Consequently, we get $\mathrm{Re} < 0.3$, which is in the low $Re$ regime. 

More details on the experimental setup and material characterization can be found in the Supplemental Information.

\subsection*{Numerical method}
In this work together with our previous studies~\cite{Lailai_2019,Lailai_2020}, we have identified and investigated an elasto-electro-hydrodynamic problem that integrates the elastohydrodynamics of flexible filaments in viscous fluids and the electrohydrodynamics of a dielectric particle in dielectric solvents. 
The numerical method adopted here closely resembles that described in the appendix of Ref.~\cite{Lailai_2020} despite two differences: first, two filaments are attached to the particle here compared to one in Ref.~\cite{Lailai_2020}; second, full 3D motion of the Quincke swimmer is pursued here, in contrast to the constrained planar motion~\cite{Lailai_2020}. 

As in Refs.~\cite{Lailai_2019,Lailai_2020}, we do not consider hydrodynamic interactions among the two filaments or those between the particle and the filaments. 
We use the semi-implicit backward Euler scheme to time-march the nonlinear governing equations in a fully coupled fashion, hence solving for the translational and rotational velocities of the particle, the induced dipole, and the instantaneous profile of the filament simultaneously. 
This in-house solver has been cross-validated against another finite-element-method (FEM) solver developed in the framework of the commercial package COMSOL Multiphysics (I-Math, Singapore). 
Before the cross-validation, the FEM solver was first validated against the numerical implementation for elasto-hydrodynamics of filaments~\cite{tornberg} and our in-house solver for the Quincke swimmers with one tail~\cite{Lailai_2019,Lailai_2020}.

\clearpage
\renewcommand{\theequation}{S\arabic{equation}}
\renewcommand{\thefigure}{S\arabic{figure}}
\renewcommand{\thetable}{S\arabic{table}}
\setcounter{equation}{0}
\setcounter{figure}{0}
\setcounter{table}{0}

\onecolumngrid
\begin{center}
    {\LARGE Supplemental Information for `Low Reynolds number, bi-flagellated Quincke swimmers with multiple forms of motion'} 
\end{center}
\vspace{5mm}

\twocolumngrid


\section{Some theoretical details on Quincke rotation}
\label{sec:QuinckeTheory}

Detailed theories on Quincke rotation of a sphere can be found in Refs.~\cite{Tsebers_1980,Jones_1984,Cebers_2001}. 
In this section, we briefly go through the essential steps to derive Eq.~(\ref{eq:AngularSpeed}) and Eq.~(\ref{eq:T_Q}) in the main text. 

When a uniform external electric field $\vec{E}_0$ is applied on a sphere (permittivity $\varepsilon_\text{s}$, electrical conductivity $\sigma_\text{s}$) with radius $R$ submerged in a liquid (permittivity $\varepsilon_\text{l}$, electrical conductivity $\sigma_\text{l}$), the electric field in space can be represented as $\vec{E} = -\nabla \varphi$, where a scalar field $\varphi$ satisfies\footnote{This is true when the system reaches a steady-state, and approximately true when the dipole oscillates slowly. } 
\begin{equation}
    \nabla^2 \varphi = 0. 
    \label{eq:QR_Laplace}
\end{equation}
Here we use subscript s to represent the region inside the sphere, and subscript l to represent the region outside the sphere, occupied by the liquid.  
The solutions to Eq.~(\ref{eq:QR_Laplace}) are 
\begin{align}
    \varphi_\text{l} &= -\vec{E}_0 \cdot \vec{r} + \frac{(\vec{A}_0+\vec{A}_1) \cdot \vec{r}}{r^3} \\
    \varphi_\text{s} &= (\vec{B}_0+\vec{B}_1) \cdot \vec{r}, 
\end{align}
where $\vec{A}_0$ and $\vec{B}_0$ are related to the bound charges, and $\vec{A}_1$ and $\vec{B}_1$ correspond to the free charges. 
$\varphi_\text{l}$ and $\varphi_\text{s}$ satisfy three boundary conditions at the sphere surface: 
\begin{itemize}
    \item Continuity of the scalar potential: 
        \begin{equation}
            \varphi_\text{l}(R) = \varphi_\text{s}(R) 
            \label{eq:BC_1}
        \end{equation}
    \item Gauss's law: 
        \begin{equation}
            \varepsilon_\text{l} \frac{\partial \varphi_\text{l}}{\partial r} \bigg|_{r = R} - \varepsilon_\text{s} \frac{\partial \varphi_\text{s}}{\partial r} \bigg|_{r = R} = \sigma_\text{f}
            \label{eq:BC_2}
        \end{equation}
    \item The continuity equation for the free charges:  
        \begin{equation}
            \frac{\partial \sigma_\text{f}}{\partial t} = - \omega \frac{\partial \sigma_\text{f}}{\partial \phi} - \sigma_\text{l} \vec{E}_\text{l}(R) \cdot \hat{r}  + \sigma_\text{s} \vec{E}_\text{s}(R) \cdot \hat{r}, 
            \label{eq:BC_3}
        \end{equation}
\end{itemize}
where $\sigma_\text{f}$ is the surface density of free charges, $\phi$ is the azimuthal angle, $\omega = \text{d}\phi / \text{d}t$ is the angular speed of the sphere, and $\hat{r}$ is the unit vector along the radial direction. 

Applying these conditions, we obtain 
\begin{equation}
    \begin{split}
        \vec{A}_0 &= \frac{\varepsilon_\text{s}-\varepsilon_\text{l}}{\varepsilon_\text{s}+2\varepsilon_\text{l}} R^3 \vec{E}_0 \\
        \vec{B}_0 &= -\frac{3\varepsilon_\text{l}}{\varepsilon_\text{s}+2\varepsilon_\text{l}} \vec{E}_0. 
    \end{split}
    \label{eq:AandB_x}
\end{equation}
They are both time independent, because bound charges have a very short response time to the variation in the external field, so they are not affected by any rotation of the sphere. 
Consequently, the angle between the induced dipole and $\vec{E}_0$ is either $0^\circ$ ($\varepsilon_\text{s} > \varepsilon_\text{l}$) or $180^\circ$ ($\varepsilon_\text{s} < \varepsilon_\text{l}$), so it applies zero torque on the sphere. 

On the contrary, the free charges have a finite relaxation time, so that $\vec{A}_1$ follows 
\begin{equation}
    \frac{d \vec{A}_1}{d t} = \vec{\omega} \times \vec{A}_1 - \frac{1}{\tau} (\vec{A}_1 - \vec{A}_{10}), 
    \label{eq:A1}
\end{equation}
where 
\begin{equation}
    \tau \equiv \frac{2\varepsilon_\text{l}+\varepsilon_\text{s}}{2\sigma_\text{l}+\sigma_\text{s}}, 
    \label{eq:RelaxationTime}
\end{equation}
is the relaxation time and 
\begin{equation}
    \vec{A}_{10} = \frac{3(\varepsilon_\text{l} \sigma_\text{s}-\varepsilon_\text{s} \sigma_\text{l})}{(2\varepsilon_\text{l}+\varepsilon_\text{s}) (2\sigma_\text{l}+\sigma_\text{s})} R^3 \vec{E}_0. 
\end{equation}
Substituting in $\vec{p} =  4\pi \varepsilon_\text{s} \vec{A}_1$, we get the dynamic equation of the dipole 
\begin{equation}
    \frac{d \vec{p}}{dt} = \vec{\omega} \times \vec{p} - \frac{1}{\tau}(\vec{p} - \vec{p}_0), 
    \label{eq:p}
\end{equation}
where 
\begin{equation}
    \begin{split}
        \vec{p}_0 & = \frac{9 \varepsilon_\text{l} (\varepsilon_\text{l} \sigma_\text{s}-\varepsilon_\text{s} \sigma_\text{l}) }{ (2\varepsilon_\text{l}+\varepsilon_\text{s}) (2\sigma_\text{l}+\sigma_\text{s}) } \frac{4}{3} \pi R^3 \vec{E}_0 \\
        & \equiv \chi_0 {\frac{4}{3} \pi R^3 \vec{E}_0}
    \end{split}
    \label{eq:p_0}
\end{equation} 
is the dipole caused by the free charges when the sphere remains stationary. 
As we can see, $\vec{p}_0$ is parallel to $\vec{E}_0$ for $\chi_0 > 0$ and $\vec{p}_0$ is anti-parallel to $\vec{E}_0$ for $\chi_0 < 0$. 

For a sphere at low Reynolds number $Re$, the electric torque on the dipole $\vec{\Gamma}_\text{Q} = \vec{p} \times \vec{E}_0$ and the viscous torque $\vec{\Gamma}_\mu = - 8 \pi \mu R^3 \vec{\omega}$, where $\mu$ is the viscosity of the ambient liquid, always balance each other: $\vec{\Gamma}_\text{Q}+\vec{\Gamma}_\mu = 0$. 
Combining this with Eq.~(\ref{eq:p}) and assuming that the electric field points toward $+x$, the set of equations that governs the rotation of the sphere is   
\begin{subequations}
\begin{align}
    & \dot{p}_x = -\omega p_y - (p_x - p_0)/\tau, \label{eq:DE_original_1} \\
    & \dot{p}_y = \omega p_x - p_y/\tau, \label{eq:DE_original_2} \\
    & 0 = -E_0 p_y - 8 \pi \mu R^3 \omega. \label{eq:DE_original_3}
\end{align}
\label{eq:DE_original}%
\end{subequations}
When the system reaches a steady state, $\dot{p}_x = \dot{p}_y = 0$, so Eqs.~(\ref{eq:DE_original}) lead to Eq.~(\ref{eq:AngularSpeed}) in the main text
\begin{equation}
    \omega = \frac{1}{\tau} \sqrt{\left( \frac{E_0}{E_\text{c}} \right)^2-1}, 
\end{equation}
where the threshold electric field is 
\begin{equation}
    E_\text{c} = \sqrt{-\frac{6 \mu}{\chi_0 \tau}}. 
    \label{eq:E_c}
\end{equation}
It is meaningful ($E_\text{c}$ is real) only when $\varepsilon_\text{l}/\sigma_\text{l} \le \varepsilon_\text{s}/\sigma_\text{s}$ so that $\chi_0 < 0$ according to Eq.~(\ref{eq:p_0}). 

Furthermore, the electric torque applied on the sphere is $\Gamma_\text{Q} = -E_0 p_y = -E_0 p_0 \dfrac{\omega \tau}{1+(\omega\tau)^2}$. 
After replacing $p_0$ with $p_0 = -\dfrac{8\pi \mu R^3 E_0}{E_\text{c}^2 \tau}$, which is obtained by combining Eq.~(\ref{eq:p_0}) and Eq.~(\ref{eq:E_c}), we get Eq.~(\ref{eq:T_Q}) in the main text
\begin{equation}
    \Gamma_\text{Q} = 8\pi \mu R^3 \left( \frac{E_0}{E_c} \right)^2 \frac{\omega}{1+(\omega \tau)^2}.
\end{equation} 
Note that in the main text, we use symbol $E$ to represent $E_0$ for simplicity.

\section{Characterize properties of liquids and swimmers}

\subsection{Electrical properties of the liquid}
\begin{figure}
    \begin{center}
        \includegraphics[scale=0.9]{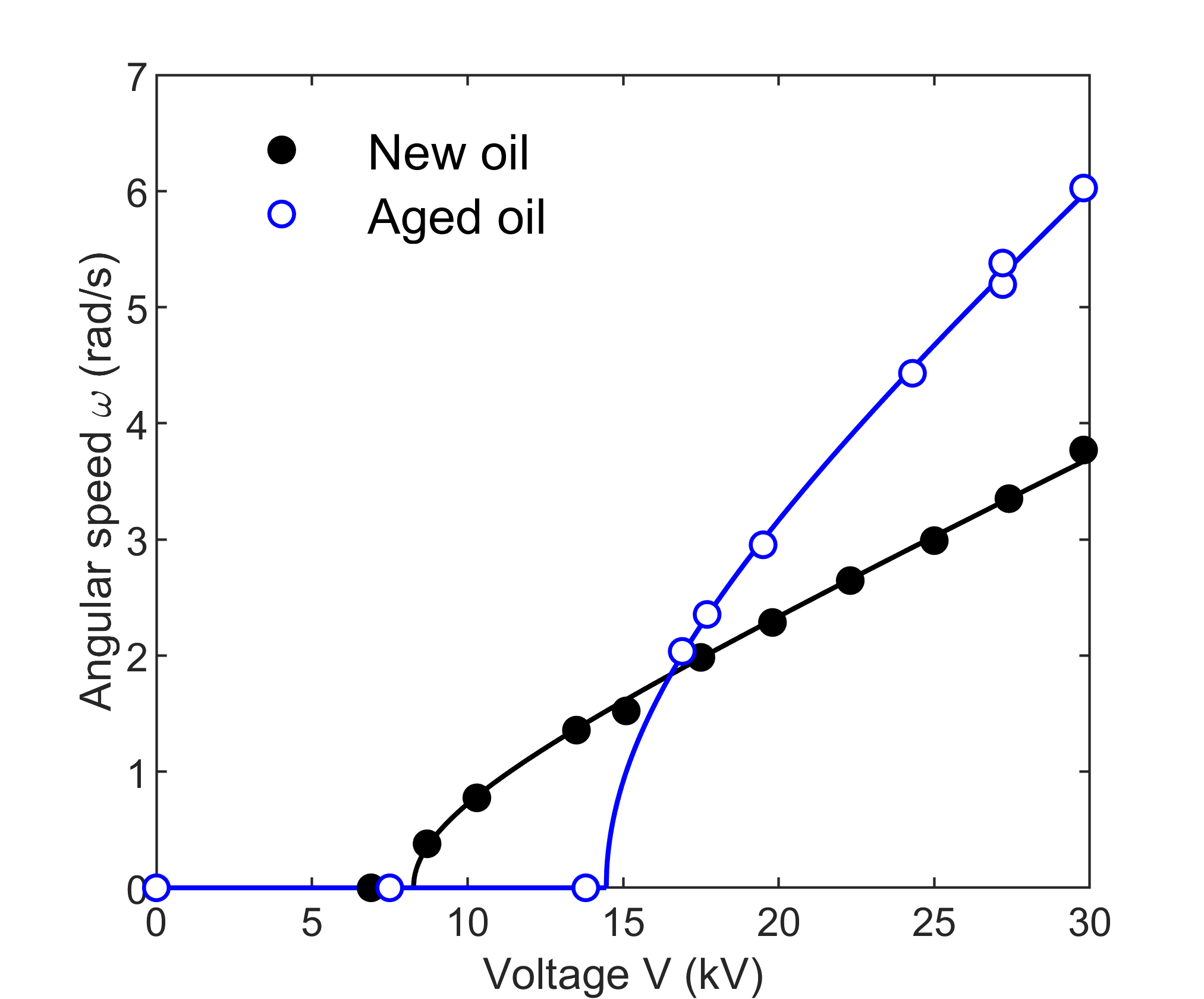}
    \end{center}
\caption{ Measure oil properties exploiting Quincke rotation of spheres. The ``New oil'' and ``Aged oil'' correspond to ``Castor oil + FB olive oil, new'' and ``Castor oil + FB olive oil, aged, Expt 1'' in Table~\ref{tab:Oil}, respectively. The solid lines show the best fits with Eq.~(\ref{eq:AngularSpeed_V}). The error bars are smaller than the size of the labels so they are not shown. }
\label{Oil}
\end{figure}

\begin{table*}
    \centering
    \begin{tabular}{ |c|c|c|c|c| } 
    \hline
    Oil composite & $E_\text{c}$ (kV/m) & ~$\tau$ (s)~ & ~$\varepsilon_\text{l} / \varepsilon_0$~ & $\sigma_\text{l}$ (S/m) \\
    \hline
    Castor oil + NF olive oil, Expt 1 (new) & 234 & 0.77 & 2.39 & $0.41 \times 10^{-10}$ \\ 
    Castor oil + NF olive oil, Expt 2 & 284 & 0.55 & 2.14 & $0.54 \times 10^{-10}$ \\ 
    Castor oil + NF olive oil, Expt 3 & 313 & 0.41 & 2.79 & $0.86 \times 10^{-10}$ \\ 
    \hline
    Castor oil + FB olive oil, new & 212 & 0.94 & 2.40 & $0.34 \times 10^{-10}$ \\ 
    Castor oil + FB olive oil, aged, Expt 1 & 371 & 0.30 & 2.56 & $1.10 \times 10^{-10}$ \\ 
    Castor oil + FB olive oil, aged, Expt 2 & 381 & 0.30 & 2.27 & $1.03 \times 10^{-10}$ \\ 
    Castor oil + FB olive oil, aged, Expt 3 & 391 & 0.26 & 2.84 & $1.36 \times 10^{-10}$ \\ 
    Castor oil + FB olive oil, aged, Expt 4 & 385 & 0.27 & 2.88 & $1.34 \times 10^{-10}$ \\ 
    \hline
    \end{tabular}
    \caption{Electrical properties of some oil mixtures used in the experiments. The first three rows show how a newly prepared oil aged after several experiments. The last four rows provide the electrical properties of the oil that we used to take the data in the main text. }
    \label{tab:Oil}
\end{table*}

We exploited Quincke rotation of HDPE spheres (with no fiber attached) to measure the electrical properties of the liquid. 
We applied different voltages $V$ and measured the angular speed $\omega$ of the sphere when it reached a steady unidirectional rotation. 
The experimental results with two different liquids are shown in Fig.~\ref{Oil}, and the data agree well with the theoretical prediction 
\begin{equation}
    \begin{aligned}
        \omega &= 0 & (&V < V_\text{c})  \\
        \omega &= \frac{1}{\tau} \sqrt{\left( \frac{V}{V_\text{c}} \right)^2-1} & (&V \ge V_\text{c})
    \end{aligned}
    \label{eq:AngularSpeed_V}
\end{equation}
where $V_\text{c}$ is the threshold voltage and $\tau$ is the relaxation time (Eq.~(\ref{eq:RelaxationTime})). 
For each liquid used in the experiments, we measured the corresponding $\omega(V)$ curve, fit the data to Eq.~(\ref{eq:AngularSpeed_V}), and obtained the two fitting parameters $V_\text{c}$ and $\tau$. 
We then calculated the threshold electric field $E_\text{c} = V_\text{c}/h$, where $h = 3.9$~cm was the distance between the inner surfaces of the copper plates. 
We list $E_\text{c}$ and $\tau$ of some oil mixtures used in the experiments in Table~\ref{tab:Oil}.  

With $E_\text{c}$ and $\tau$, we calculated the electric permittivity $\varepsilon_\text{l}$ and conductivity $\sigma_\text{l}$ of the liquid. 
The conductivity of the HDPE sphere $\sigma_\text{s}$ was at least five orders of magnitude smaller than $\sigma_\text{l}$, so $\sigma_\text{s} = 0$ is a good approximation. 
The permittivity of the solid sphere was $\varepsilon_\text{s} = 2.4 \varepsilon_0$, where $\varepsilon_0 = 8.85 \times 10^{-12}$~F/m. 
Substituting $\varepsilon_\text{s}$ and $\sigma_\text{s}$ into Eq.~(\ref{eq:RelaxationTime}) and Eq.~(\ref{eq:E_c}), and rearranging the terms, we obtain  
\begin{subequations}
    \begin{align}
        & 2\varepsilon_\text{l} + \varepsilon_\text{s} = 2 \tau \sigma_\text{l} \\ 
        & \varepsilon_\text{l} \varepsilon_\text{s} = \frac{8\mu \sigma_\text{l}}{3 E_\text{c}^2}. 
    \end{align}
\label{eq:elecProperties}%
\end{subequations}
For the oil mixtures listed in Table~\ref{tab:Oil}, we calculated their $\varepsilon_\text{l}$ and $\sigma_\text{l}$ using Eqs.~(\ref{eq:elecProperties}). 
In general, the permittivity of an oil remained invariant approximately, and the conductivity increased as it aged during the experiments. 


We used two different types of olive oil in the castor oil-olive oil mixtures. 
One was food grade olive oil (Filippo Berio), the other was NF grade oil (Spectrum), which had a higher purity. 
The castor oil (Alfa Aesar) we used was always the same.
The newly prepared oil mixtures using the two different olive oils (row 1 and row 4 in Table~\ref{tab:Oil}) had very close $\varepsilon_\text{l}$ and slightly different $\sigma_\text{l}$. 
As the experiments went on (row 1 to row 3 in Table~\ref{tab:Oil}), the oil aged due to copper ions dissolved into the oil\footnote{Visually, the oil slowly turned from a yellow color to a light green color.} and possibly other reasons. 
The consequence was that the conductivity increased from experiment to experiment. 
To control this aging effect, we mixed some newly made mixed oil into the aged oil before each experiment, to keep both $E_\text{c}$ and $\tau$ approximately invariant throughout the experiments (row 5 to row 8 in Table~\ref{tab:Oil}).

\subsection{Measure diameter of the tails}
Since the bending stiffness of a fiber is proportional to the fourth power of its radius, it is vital to measure the radii of the fibers that we used to make the tails of the swimmers. 
In the experiments, we used No.8 and No.10 surgical sutures, and we measured their diameters using a microscope (Nikon Ti-E inverted with PFS). 
We took images at different locations of each fiber and confirmed that its diameter was uniform. 
The mean diameter of the No.8 surgical suture (S\&T) was $49.5 \pm 0.6~\mu$m, and the mean diameter of the No.10 surgical suture (DemeTech) was $27.3 \pm 0.5~\mu$m.

\subsection{Measure bending stiffness of the tails}
\begin{figure}
    \begin{center}
        \includegraphics[width = 0.5\textwidth]{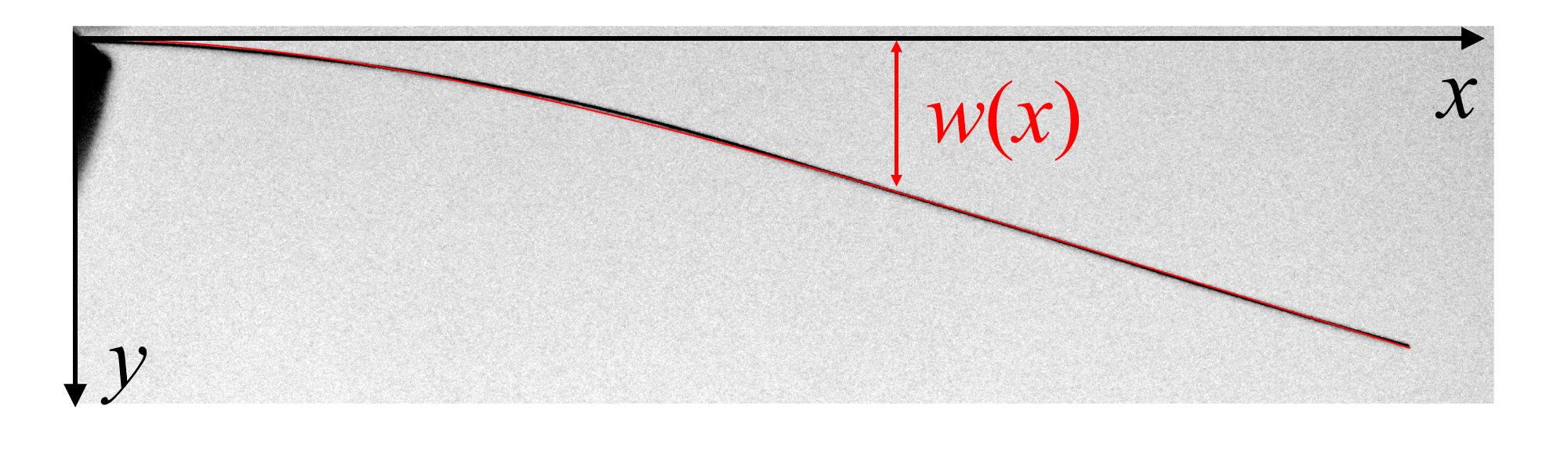}
    \end{center}
\caption{ Fit the shape of the fiber with Eq.~(\ref{eq:bending_fit}). The fiber is black in the image, and the fit is shown by the red curve.  }
\label{FiberStiffness}
\end{figure}

The bending stiffness of the nylon fibers that were used to make swimmer tails were measured by hanging one end of the fiber horizontally and keeping the other end free, as shown in Fig.~\ref{FiberStiffness}. 
The shape of the fiber is described by 
\begin{equation}
    w(x) = \frac{q x^2 (x^2 - 4Lx + 6L^2)}{24YI_\text{f}}, 
    \label{eq:bending}
\end{equation}
where $w(x)$ is the vertical displacement of the fiber at position $x$, $q$ is the linear density of the body force, $L$ is the total horizontal length of the fiber, $Y$ is the Young's modulus, and $I_\text{f}$ is the second moment of area of the fiber \cite{Landau_book}. 
Substitute in $q = \pi a^2 \rho g$ and $I_\text{f} = \pi a^4 / 4$, where $a = 24.8~\mu$m is the fiber radius, $\rho = 1.15$~g/cm$^3$ is the fiber density, and $g$ is the gravitational acceleration, we obtain 
\begin{equation}
    w(x) = \left( \frac{\rho g}{6Ya^2} \right) x^2 (6L^2 - 4Lx + x^2). 
    \label{eq:bending_fit}
\end{equation}
By fitting the shape of the fiber bent under gravity with Eq.~(\ref{eq:bending_fit}), we obtain its Young's modulus $Y = 2.7 \pm 0.5$~GPa, which agrees with the numbers given for Nylon 6/6 in the literature. 
Consequently, the bending stiffness of the fiber is $\kappa = Y I_\text{f} = 7.9 \times 10^{-10}$~Nm$^2$.

\subsection{Geometry of the swimmer}

\begin{figure*}
    \begin{center}
        \includegraphics[scale=0.5]{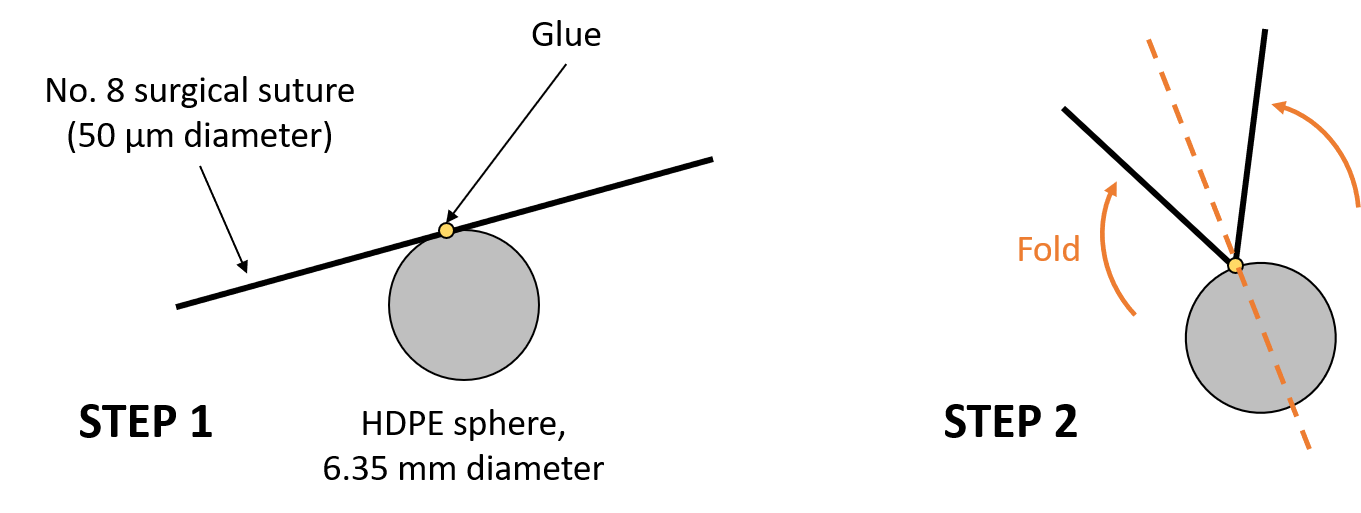}
    \end{center}
\caption{ Preparation of the swimmer. Step 1: Cut a piece of nylon fiber out of No.8 surgical suture, and attach its center to the HDPE sphere using a small drop of glue. Step 2: Fold the fibers up to a certain angle, press the bottom of the fiber until it deforms plastically.  }
\label{SwimmerPreparation}
\end{figure*}

\begin{figure*}
    \begin{center}
        \includegraphics[scale = 0.7]{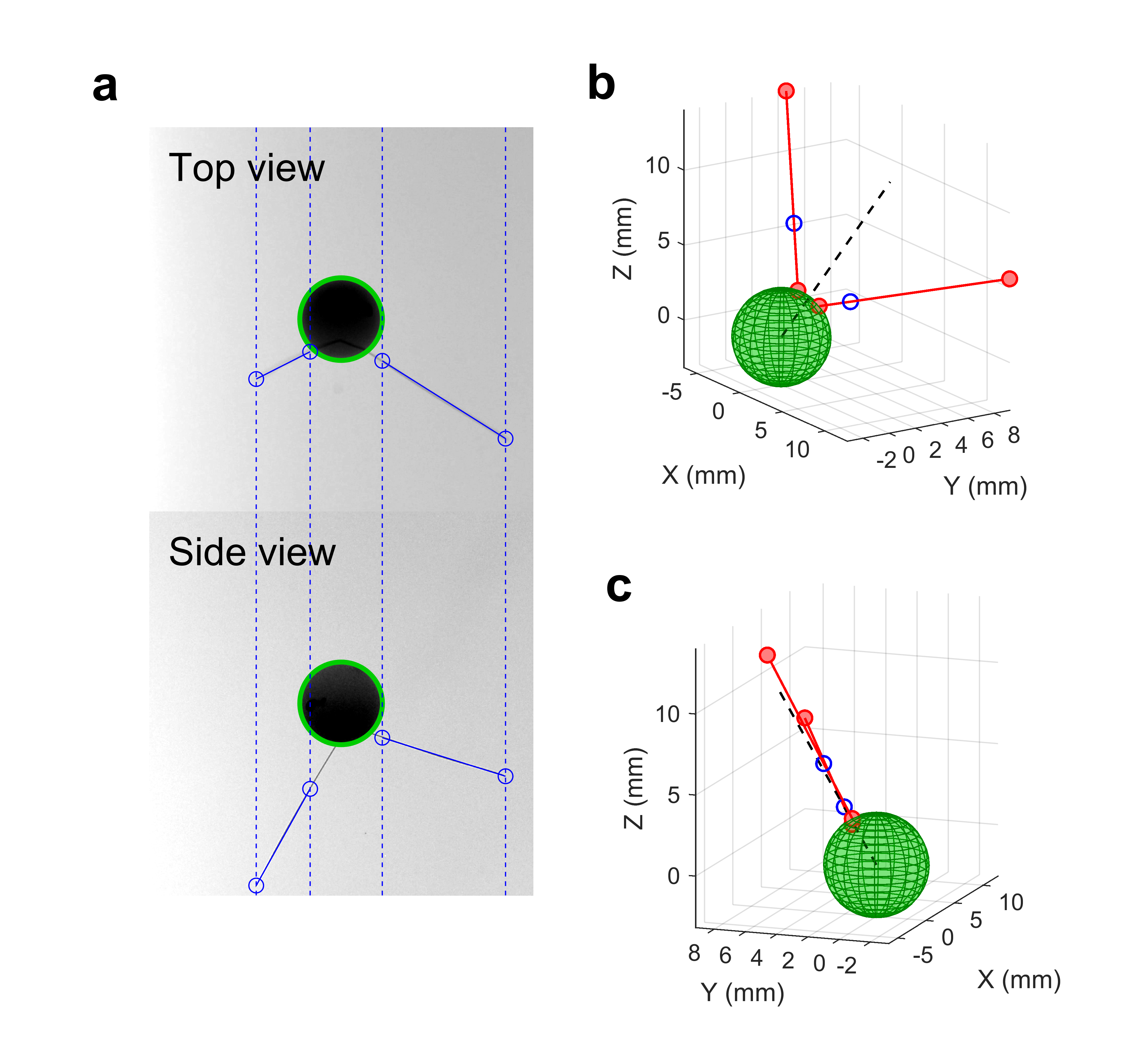}
    \end{center}
\caption{ Measure the geometric parameters of the swimmer. 
(a) Top view and side view of the swimmer resized, cropped, and shown next to each other. The edge of the sphere is highlighted by the green circle. The blue circles are the points on the tails chosen by hand. The two points on each vertical dashed line are supposed to be the same point on the swimmer in 3D. 
(b,c) Reconstructed three dimensional model of the swimmer viewed from two perspectives. The red circles label the ends and bases of the tails. The blue circles correspond to those labeled in (a) (at the end of the tails, the blue circles are covered by the red circles). The dashed black line shows the axis of symmetry, which goes through the center of the sphere and the middle point of the tail bases. }
\label{InitialShape}
\end{figure*}

The preparation process of the swimmer is sketched in Fig.~\ref{SwimmerPreparation}. 
To measure the geometric parameters of the swimmer such as the tail length $L$ and the tail angle $\theta$, for each swimmer, we performed at least one experiment with no field applied. 
A typical stationary swimmer is shown in Fig.~\ref{InitialShape}. 
Its configuration in three dimensions was reconstructed following these steps: \\
1. Resize the top view and the side view images taken at the same time so that they have the same length scale (mm/pixel). \\
2. Crop the images so that the center of the sphere is at the center of each image, then display them side by side as shown in Fig.~\ref{InitialShape}(a). \\
3. Locate two points on each tail manually, with one point being the end of the tail. These selected points are labeled by the blue circles in Fig.~\ref{InitialShape}(a). \\
4. With the $x$, $y$, and $z$ coordinates of these selected points, we located the positions and orientations of the tails, and found where they intersected with the sphere surface. These intersections are the bases of the tails. The reconstructed swimmer in three dimensions is shown in Fig.~\ref{InitialShape}(b,c), with the tips and bases of the tails labeled by solid red circles. \\
5. The axis of symmetry is found by connecting the center of the sphere (at (0,0,0)) and the middle point between the bases of the tails, as shown by the dashed black lines in Fig.~\ref{InitialShape}(b,c). 

With the reconstructed three-dimensional configuration, we calculated the tail lengths $L$, the tail angle $\theta$, and the distance between the tail bases $\delta_\text{t}$. 
If the unit vectors along the two tails are $\hat{r}_1$ and $\hat{r}_2$, respectively, the tail angle is defined as 
\begin{equation}
    \theta = \frac{1}{2}\text{arccos}(\hat{r}_1 \cdot \hat{r}_2). 
    \label{eq:TailAngle}
\end{equation}
The geometric parameters of the swimmers shown in Fig.~\ref{StateDiagram} in the main text are listed in Table~\ref{tab:Swimmer}. 
\begin{table}
    \centering
    \begin{tabular}{ |c|c|c|c| } 
    \hline
    ~$\theta$ ($^\circ$)~ & ~$L_1$ (mm)~ & ~$L_2$ (mm)~ & ~$\delta_\text{t}$ (mm)~ \\
    \hline
    17.4 & 13.9 & 13.3 & 2.4 \\
    39.9 & 13.8 & 13.1 & 2.4 \\
    43.7 & 14.2 & 13.4 & 1.6 \\
    46.9 & 14.9 & 13.9 & 1.8 \\
    49.5 & 13.8 & 13.4 & 2.2 \\
    57.5 & 13.6 & 12.8 & 2.1 \\
    79.9 & 15.0 & 13.3 & 0.3 \\
    \hline
    \end{tabular}
    \caption{Geometric parameters of the swimmers shown in Fig.~\ref{StateDiagram} in the main text. $L_1$ is the length of the longer tail and $L_2$ is the length of the shorter tail. }
    \label{tab:Swimmer}
\end{table}
The average tail length was $L = 13.7 \pm 0.6$~mm, and the distance between the bases was $\delta_\text{t} = 2.1 \pm 0.3$~mm (after removing the outlier). 
A perfectly symmetric swimmer should have $L_1 = L_2$, but in practice, we had a small asymmetry in the tail length, where $|L_1 - L_2| = 0.8 \pm 0.4$~mm.

\section{Measure translational and rotational speed of the sphere}

\begin{figure*}
    \begin{center}
        \includegraphics[scale = 1]{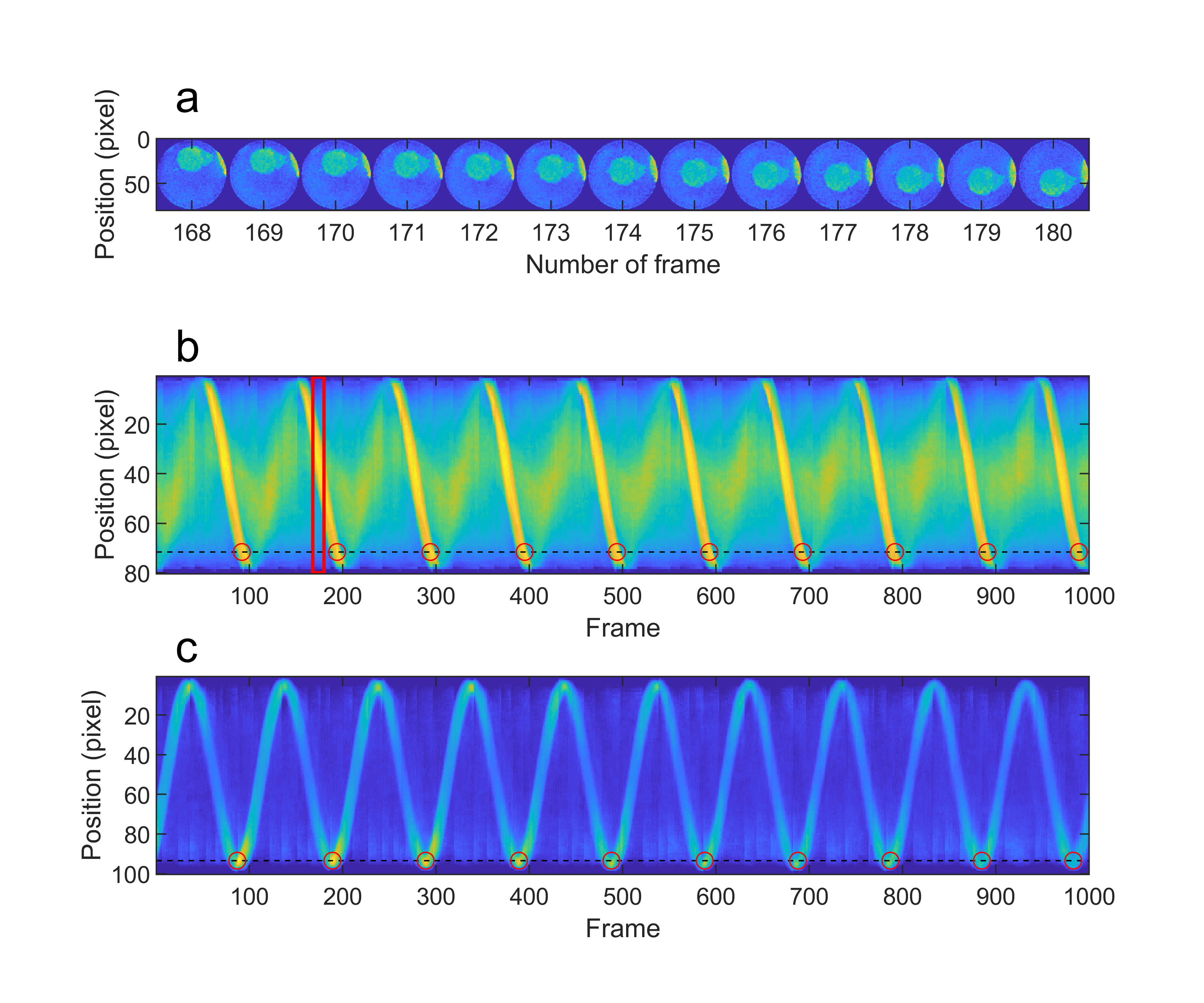}
    \end{center}
\caption{ Measure the period of rotation when a sphere with colored labels on the surface rotated continuously toward one direction. 
(a) A series of cropped images that contain the sphere. 
(b,c) Kymographs showing how the surface brightness of the sphere changes with time from (b) top view and (c) side view. The frames shown in (a) are highlighted by the red box in (b). 
The dashed black lines show where the peaks were detected, and the located peaks are labelled by the red circles. }
\label{AngSpeedImgSeries}
\end{figure*}

\begin{figure*}
    \begin{center}
        \includegraphics[scale = 0.55]{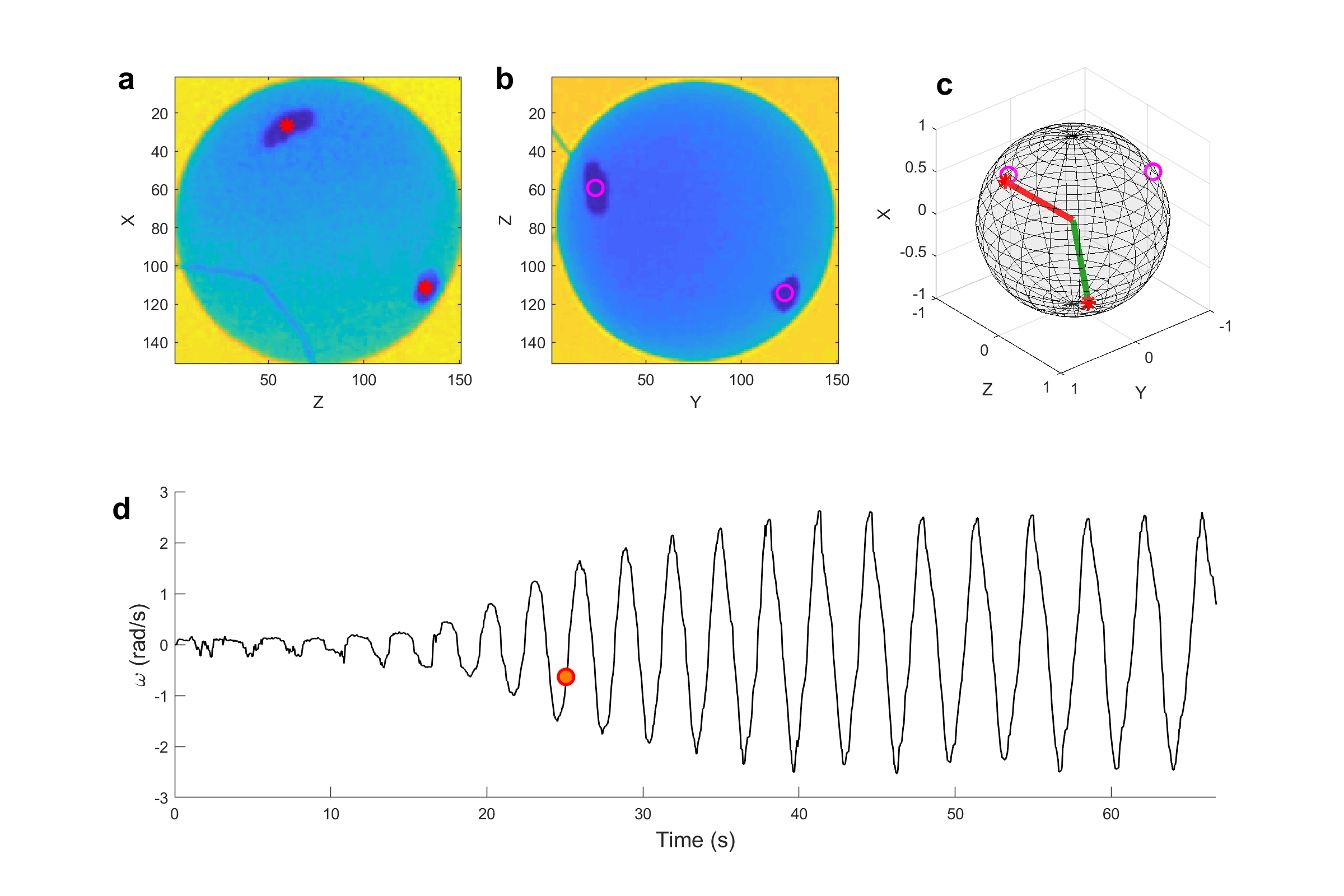}
    \end{center}
\caption{ Measure the instantaneous angular speed of the swimmer. 
(a) Side view and (b) top view of the cropped images with the sphere, with the colored spots identified and labeled by red asterisks in (a) and magenta circles in (b).  
(c) Locations of the spots on the surface of the sphere. The labels are the same as in (a) and (b). The red and green lines show the two vectors that we used to measure the three Euler angles. 
(c) Instantaneous angular speed $\omega$ as a function of time. The red dot labels the current frame shown in (a)-(c). 
The swimmer in the video had a particle radius $R = 3.18$~mm, tail radius $a = 24.8~\mu$m, tail length $13.8$~mm, and tail angle $\theta = 43.7^\circ$. The applied electric field was $E = 1.65 E_\text{c}$. }
\label{SphereOrientationTracking}
\end{figure*}

\subsection{Measure position and velocity of the sphere}
In each experiment, we obtained two videos: one from the top camera and the other from the side camera. 
The top camera captured the $x$-$y$ plane, and the side camera captured the $x$-$z$ plane. 
We tracked the center position of the sphere in every frame in both videos. 
Since both cameras captured the $x$ position of the sphere, we used $x(t)$ to resize the two videos so that they had the same length scale. 
Then we used the radius of the sphere ($R$ = 3.18~mm) to calculate the length scale (mm/pixel). 
With the positions $x(t)$, $y(t)$, and $z(t)$ as functions of time, we calculated the velocity of the sphere and the corresponding viscous drag, as described in the main text.

\subsection{Measure angular speed when the sphere rotates continuously}
We introduced two methods to measure the angular velocity of the sphere. 
For a sphere rotating with a constant angular speed, the method we used is illustrated in Fig.~\ref{AngSpeedImgSeries}, and the procedure is as follows: \\
1. In each frame, we cropped a square that contained the sphere. An exemplary series of such cropped images are shown in Fig.~\ref{AngSpeedImgSeries}(a). \\
2. We averaged each cropped image along the horizontal axis, so that each square became a column. \\
3. We put all the columns in a horizontal series to get the kymographs shown in Fig.~\ref{AngSpeedImgSeries}(b) (top view) and Fig.~\ref{AngSpeedImgSeries}(c) (side view). The frames shown in Fig.~\ref{AngSpeedImgSeries}(a) are highlighted by the red box in Fig.~\ref{AngSpeedImgSeries}(b). \\
4. We chose a horizontal line by hand so that in each period, one peak was located along that line. The locations of the peaks are labeled by the red circles in the figure. \\
5. The horizontal distance between adjacent red circles represents a period of the rotation. With all the peaks located, we calculated the average angular speed $\omega$ and its standard deviation.

\subsection{Measure angular speed when the sphere oscillates}
When the sphere oscillates, the method above can still provide the period of oscillation $T$, but not the instantaneous angular speed $\omega(t)$. 
To measure $\omega$, we tracked the colored spots labeled on the surface of the sphere, following the steps below: \\ 
1. In each frame, we cropped a square that contained the sphere. 
Such cropped images are shown in Fig.~\ref{SphereOrientationTracking}(a) - side view and Fig.~\ref{SphereOrientationTracking}(b) - top view. \\
2. We located the positions of the colored spots on the surface of the sphere, as labelled by the asterisks in Fig.~\ref{SphereOrientationTracking}(a) and circles in Fig.~\ref{SphereOrientationTracking}(b). \\ 
3. We calculated the locations of the spots in three dimensions, as shown in Fig.~\ref{SphereOrientationTracking}(c). The dimensions are normalized by the radius of the sphere. \\ 
4. The orientation of the sphere can be represented by the three Euler angles $\alpha$, $\beta$, and $\gamma$. To find all three angles, we need to identify at least two points on the surface of the sphere, and our rules are: \\
(1) Find spot with the largest area in the side view image, connect it with the center of the sphere shown by the red line in Fig.~\ref{SphereOrientationTracking}(c). The orientation of the red line is $(\text{sin}\alpha \cdot \text{sin}\beta,~-\text{cos}\alpha \cdot \text{sin}\beta,~\text{cos}\beta)$. \\
(2) If there is another spot detected in the side view image, connect it with the center of the sphere using a green line. If not, find the largest spot detected on the top view image, and make sure that it is not the same spot that we used in (1). Connect this point with the center of the sphere using a green line. \\
(3) The orientation of the green line with respect to the red line gives us the third angle $\gamma$. \\
5. Calculate the angular speed using 
\begin{equation}
    \omega = \text{sgn}(\dot{\beta}) \cdot \sqrt{\dot{\alpha}^2 + \dot{\beta}^2 + \dot{\gamma}^2 + 2\dot{\alpha}\dot{\gamma} \text{cos}\beta }, 
\end{equation}
where $\text{sgn}()$ is the sign function: it is 1 when $\dot{\beta} \ge 0$, and it is $-1$ when $\dot{\beta} < 0$. 

Following the steps above we obtained the curve in Fig.~\ref{SphereOrientationTracking}(d). 
With such $\omega(t)$ curves, we measured the oscillation period $T$ and the maximum angular speed $\omega_0$ shown in Fig.~\ref{AngularSpeed} of the main text.

\section{Calculating onset electric field}
\label{sec:Onset}

\begin{figure}
    \begin{center}
        \includegraphics[scale=0.7]{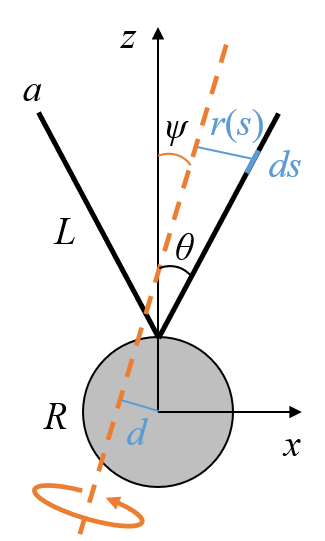}
    \end{center}
\caption{ Geometry of the swimmer. The rotation axis, which is shown by the orange dashed line, is in the same plane as the two tails of the swimmer. Its position can be described by its distance to the center of the sphere $d$ and the angle $\psi$.  }
\label{SwimmerSketch}
\end{figure}

As discussed in the main text, to calculate the onset electric fields, we only need to consider swimmers with rigid tails. 
The geometry of an exemplary swimmer is illustrated in Fig.~\ref{SwimmerSketch}. 
The question is: If the swimmer is rotating along any axis that is in plane with the two tails, which is labeled by the dashed orange line in Fig.~\ref{SwimmerSketch}, what is the total torque applied on the swimmer by the hydrodynamic resistance? 
The radius of the sphere is $R$, the tails have radius $a$ and length $L$, and the tail angle is $\theta$. 
The angle between the rotation axis and the $z$ axis is defined as $\psi$, and the distance between the rotation axis and the center of the sphere is $d$. 

\begin{figure*}
    \begin{center}
        \includegraphics[scale = 0.95]{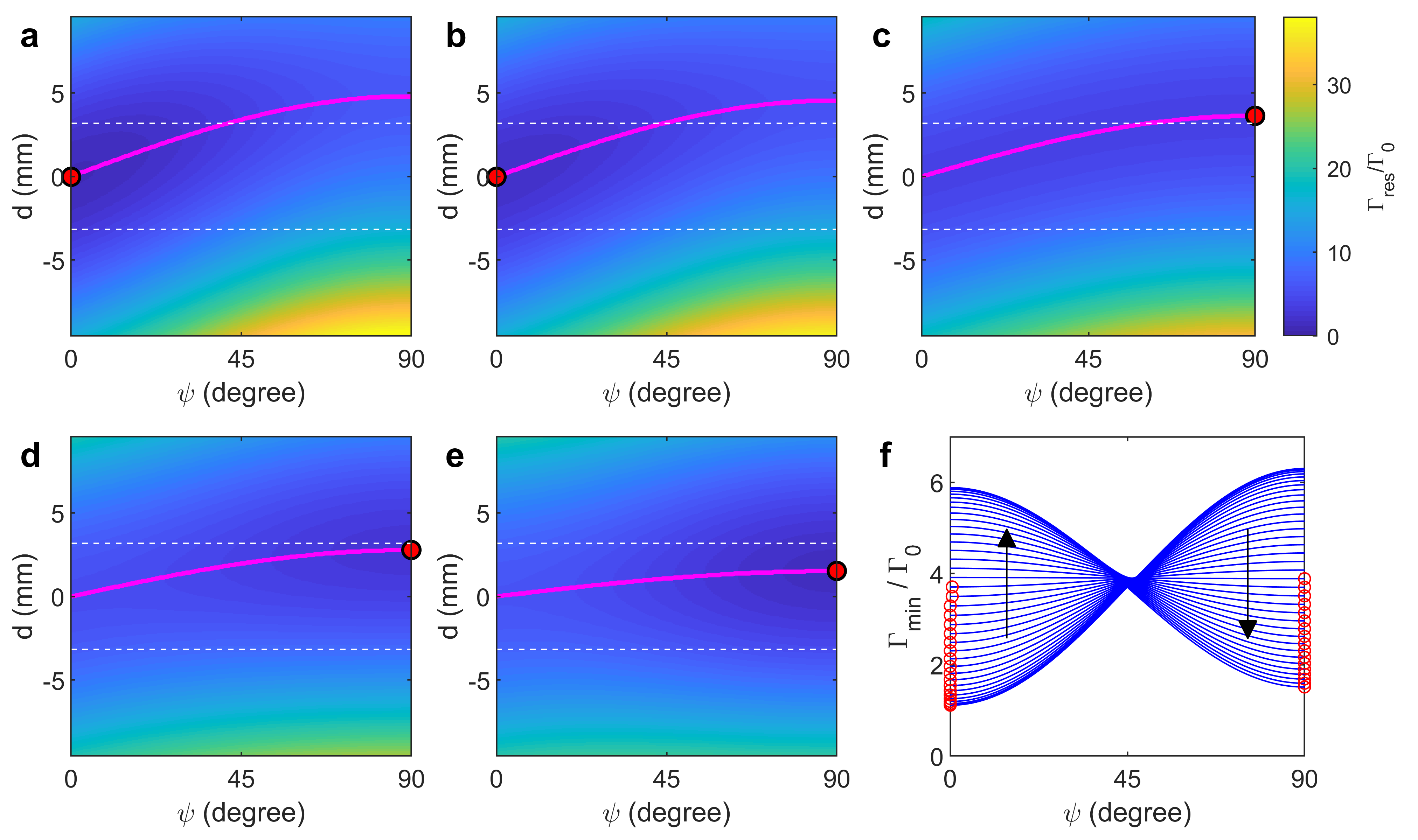}
    \end{center}
\caption{ Find the axis with the minimum torque assuming the tails are rigid. (a) to (e) show the resistance torque $\Gamma_\text{res}$ applied on these rigid swimmers at $\omega = 1$~s$^{-1}$. 
The corresponding tail angles $\theta$ are (a) $0^\circ$, (b) $22.5^\circ$, (c) $50^\circ$, (d) $67.5^\circ$, and (e) $90^\circ$. 
The magenta curves indicate minimum torque at every $\psi$, and the red circle labels the minimum torque for the whole panel. 
The dashed white lines indicate the size of the sphere. 
(f) shows $\Gamma_\text{res}$ as a function of $\psi$ along the magenta curves in the 2D maps like (a)-(e) when the swimmer has different tail angles $\theta$. The red circles label the point with minimum $\Gamma_\text{res}$ on each curve. The black arrows represent the direction of increasing $\theta$. }
\label{MinimumTorque}
\end{figure*}

The total torque $\Gamma_\text{res}$ applied on the swimmer when it rotates around the orange axis has three components: \\ 
(1) The torque due to the spinning of the sphere (about its center)
\begin{equation}
    \Gamma_\mu = - 8 \pi \mu R^3 \omega. 
    \label{eq:SI_T_nu}
\end{equation}
Since the corresponding total force is zero, this torque is invariant when we shift the axis of rotation parallelly. \\
(2) The torque due to the off-center rotation of the sphere
\begin{equation}
    \Gamma_\text{ec} = - 6 \pi \mu R d^2 \omega.  
    \label{eq:SI_T_o}
\end{equation}
(3) The torques applied on each tail $\Gamma_\text{t}$. 
When a slender body is moving with speed $v$ perpendicular to its long axis, the applied hydrodynamic drag per unit length is 
\begin{equation}
    \frac{\text{d} f_\perp(s)}{\text{d} s} \approx -\frac{4 \pi \mu v}{\text{ln}~\epsilon^{-1}}, 
    \label{eq:f_slender}
\end{equation}
where $\epsilon = a/L$ is the ratio between the fiber radius $a$ and the fiber length $L$ \cite{Stone_Chapter_2012}. 
In our experiments $a = 24.8~\mu$m and $L = 13.7$~mm, so $\epsilon = 1.8 \times 10^{-3}$ and $\text{ln}~\epsilon^{-1} \approx 6.3$. 
As shown in Fig.~\ref{SwimmerSketch}, for a small section on a tail, its position along the tail is $s$ and its distance to the axis of rotation is $r(s)$. 
The torque applied on the whole tail is thus 
\begin{equation}
    \Gamma_\text{t} = \int_0^L r(s) \text{d}f_\perp(s) = -\frac{4 \pi \mu \omega}{\text{ln}~\epsilon^{-1}} \int_0^L r^2(s) \text{d}s, 
    \label{eq:SI_T_t}
\end{equation}
where $\Gamma_\text{t}$ needs to be calculated separately for each tail ($\Gamma_{\text{t}1}$ and $\Gamma_{\text{t}2}$). 
We can define the range of $\psi$ and the sign of $d$ as $\psi \in [0^\circ,~ 90^\circ]$; $d$ is positive when the rotation axis intersects with the positive $z$ axis, and negative when it intersects with the negative $z$ axis. 
Consequently, for the tail on the right, the distance $r(s)$ is 
\begin{equation}
    r_1(s) = | (d-R \text{sin} \psi) + s \cdot \text{sin}(\theta - \psi) |, 
    \label{eq:SI_r1}
\end{equation}
and for tail on the left 
\begin{equation}
    r_2(s) = | -(d-R \text{sin} \psi) + s \cdot \text{sin}(\theta + \psi) |. 
    \label{eq:SI_r2}
\end{equation}
Combine Eq.~(\ref{eq:SI_T_nu}) to Eq.~(\ref{eq:SI_r2}), the absolute value of the total torque of hydrodynamic resistance is 
\begin{equation}
    \Gamma_\text{res} = | \Gamma_\mu + \Gamma_\text{ec} + \Gamma_\text{t1} + \Gamma_\text{t2} |. 
    \label{eq:Torque_resistance}
\end{equation}

For a given swimmer with a certain tail angle $\theta$, $\Gamma_\text{res}$ is a function of $d$ and $\psi$. 
Numerically calculated $\Gamma_\text{res}(d,\psi)$ at five different tail angles are shown in Fig.~\ref{MinimumTorque}(a-e). 
In the calculation, we set $\mu = 0.225$~Pa$\cdot$s and $\omega = 1$~s$^{-1}$. 
In each panel, the color map represents the magnitude of $\Gamma_\text{res}(d,\psi)$, and the magenta curve shows the minimum $\Gamma_\text{res}$ at each $\psi$, which we call $\Gamma_\text{min}(\psi)$. 
The red spot indicates the minimum $\Gamma_\text{res}$ on the magenta curve, and thus the whole panel. 
Fig.~\ref{MinimumTorque}(f) shows some $\Gamma_\text{min}(\psi)$ curves from $\theta = 0^\circ$ to $\theta = 90^\circ$, where $\Gamma_\text{min}$ is normalized by $\Gamma_0 = \kappa/L$. 
As $\theta$ increases, the torque at small $\psi$ increases monotonically, and at the same time, the torque at large $\psi$ decreases monotonically. 
When a threshold $\theta_\text{c} = 49.8^\circ$ is reached, $\Gamma_\text{min}(\psi)$ is almost flat and the overall minimum (red circles) shifts from $\psi = 0$ to $\psi = 90^\circ$. 

Assuming the swimmer rotates about the axis that minimizes the viscous torque, we predict that the swimmer will start rotating about the symmetric axis ($\psi = 0^\circ$) when $\theta < \theta_\text{c}$ and a perpendicular axis ($\psi = 90^\circ$) when $\theta > \theta_\text{c}$, which agrees with the experiments. 
Moreover, we obtained the onset electric field shown in Fig.~\ref{StateDiagram} in the main text by comparing the slope of the electric torque $\text{d}\Gamma_\text{Q} / \text{d}\omega$ and the slope of the resistant torque $\text{d}\Gamma_\text{res} / \text{d}\omega$ at $\omega \to 0$~s$^{-1}$. 
The transition happens when these two slopes are equal. 
According to Eq.~(\ref{eq:T_Q}) in the main text, 
\begin{equation}
    \frac{\text{d}\Gamma_\text{Q}}{\text{d} \omega} \bigg|_{\omega = 0} = 8 \pi \mu R^3 \left( \frac{E}{E_\text{c}} \right)^2. 
    \label{eq:SI_TQ_slope}
\end{equation}
For the resistant torque, we define the minimum $\Gamma_\text{res}(d,\psi)$ at each $\theta$ as $\Gamma_\text{MIN}(\theta)$, which corresponds to the red circles in Fig.~\ref{MinimumTorque}(f). 
According to Eqs.~(\ref{eq:SI_T_nu})-(\ref{eq:Torque_resistance}), $\Gamma_\text{res}$ is a linear function of $\omega$, so 
\begin{equation}
    \frac{\text{d}\Gamma_\text{MIN}(\theta,\omega)}{\text{d} \omega} \bigg|_{\omega = 0} = \frac{\Gamma_\text{MIN}(\theta, \omega = 1)}{1}. 
    \label{eq:SI_TMIN_slope}
\end{equation}
At each $\theta$, the threshold electric field is $E_\theta$, which makes
\begin{equation}
    \frac{\text{d}\Gamma_\text{Q}(E = E_\theta)}{\text{d} \omega} \bigg|_{\omega = 0} = \frac{\text{d}\Gamma_\text{MIN}(\theta,\omega)}{\text{d} \omega} \bigg|_{\omega = 0}. 
\end{equation}
Substitute in Eq.~(\ref{eq:SI_TQ_slope}) and Eq.~(\ref{eq:SI_TMIN_slope}), we obtain 
\begin{equation}
    \frac{E_\theta}{E_\text{c}} = \sqrt{ \frac{\Gamma_\text{MIN}(\theta,\omega=1)}{8\pi \mu R^3} }, 
\end{equation}
and this is the solid black curve in Fig.~\ref{StateDiagram} of the main text. 

In the calculations we assumed the distance between the tail bases $\delta_\text{t} = 0$ to keep the model simple. 
In reality, there was always some distance between the bases of the two tails, as listed in Table~\ref{tab:Swimmer}. 
The simple model works reasonably well.

\section{Other swimmers}

\subsection{Swimmer with one tail}
\begin{figure}
    \begin{center}
        \includegraphics[scale = 0.95]{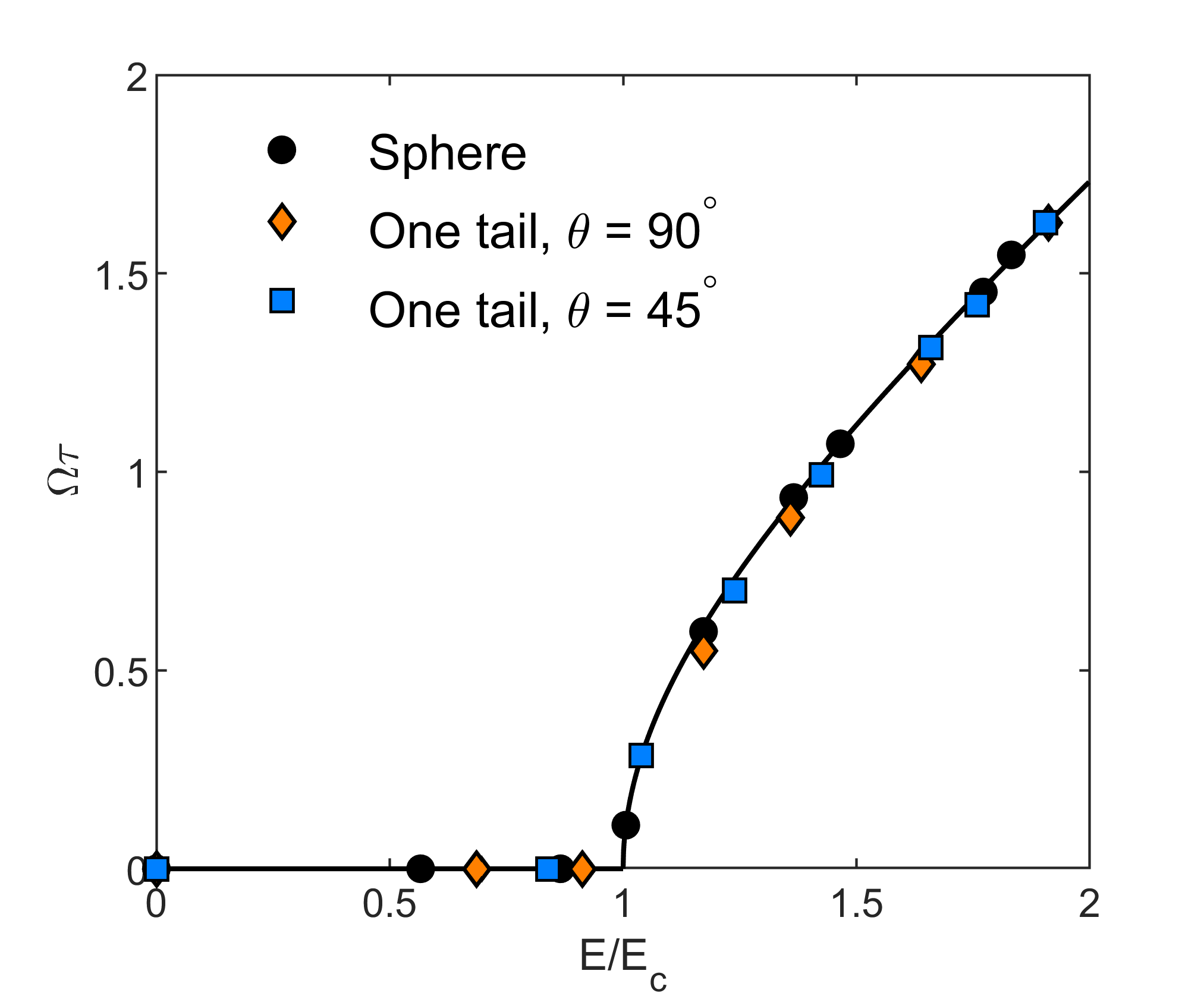}
    \end{center}
\caption{ Angular speed as a function of applied electric field for one-tail swimmers in comparison with a sphere ($R = 3.18$~mm). The two one-tail swimmers had tail angles $\theta = 45^\circ$ and $\theta = 90^\circ$, respectively. The tail length was $L = 13.4$~mm. }
\label{SI_OneTail}
\end{figure}

Numerical simulations have shown that when the swimmer has only one elastic tail, the transition from stationary to steady rotation changes from a pitchfork bifurcation to a Hopf bifurcation as $E$ increases, and an undulating mode results in a non-reciprocal motion that allows the swimmer to propel itself \cite{Lailai_2019,Lailai_2020}.
Experimentally, however, we found that without any confinement, this undulating mode was not preferred by the swimmer. 
The swimmers in our experiments had the same spherical head as the ones discussed in the main text, and their tails were made of the same nylon fiber (24.8~$\mu$m in radius). 
Similar to the bi-flagellated swimmers, the tail angle $\theta$ is defined as the angle between the tail and the line that goes through the base of the tail and the center of the sphere.  
As the external electric field $\vec{E}$ increased, the swimmer tended to rotate so that its tail was perpendicular to $\vec{E}$ instead of parallel to it. 
Then the swimmer rotated about an axis that was roughly along the line that connected the center of the sphere and the tip of the tail regardless of $\theta$, which caused limited deformation of the tail. 
Consequently, the tail did not apply significant torque on the sphere, and the angular speed of the rotating one-tail swimmer was very close to that of a bare sphere at the same $E/E_\text{c}$, as shown in Fig.~\ref{SI_OneTail}. 
The propulsive force $F_\text{t}$ was measured with the same method used for the bi-flagellated swimmers, but it was not significantly above the background drift speed obtained with a bare sphere as expected.

\subsection{Smaller swimmers}

\begin{figure}
    \begin{center}
        \includegraphics[scale = 0.95]{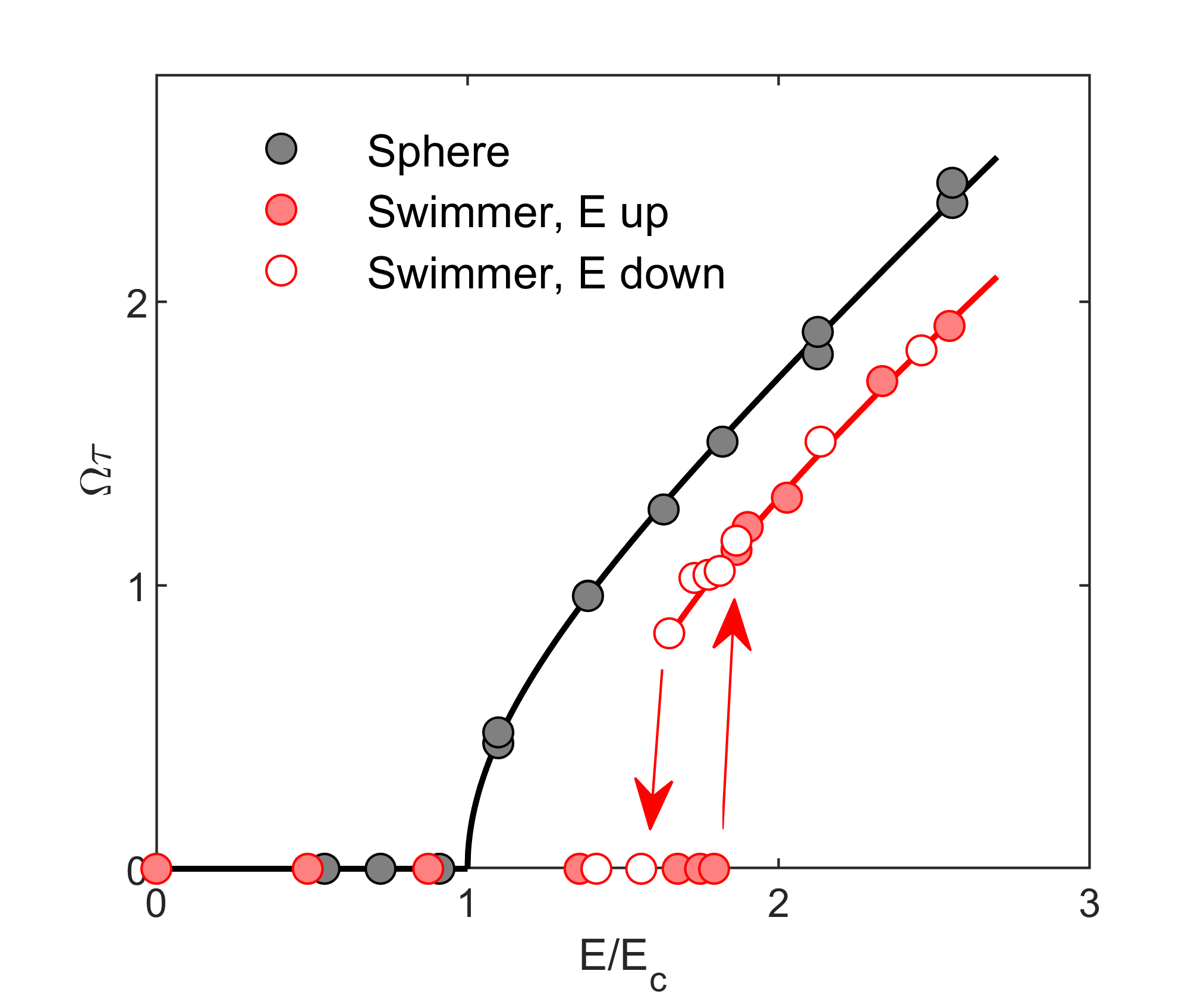}
    \end{center}
\caption{ Angular speed as a function of applied electric field for a small swimmer with tail angle $\theta = 43^\circ$, tail length $L = 9.3$~mm, and spherical head radius $R = 1.59$~mm. The arrows highlight the hysteresis when $E$ increased and decreased. }
\label{SI_SmallSwimmer_AngularSpeed}
\end{figure}

\begin{figure}
    \begin{center}
        \includegraphics[scale = 0.55]{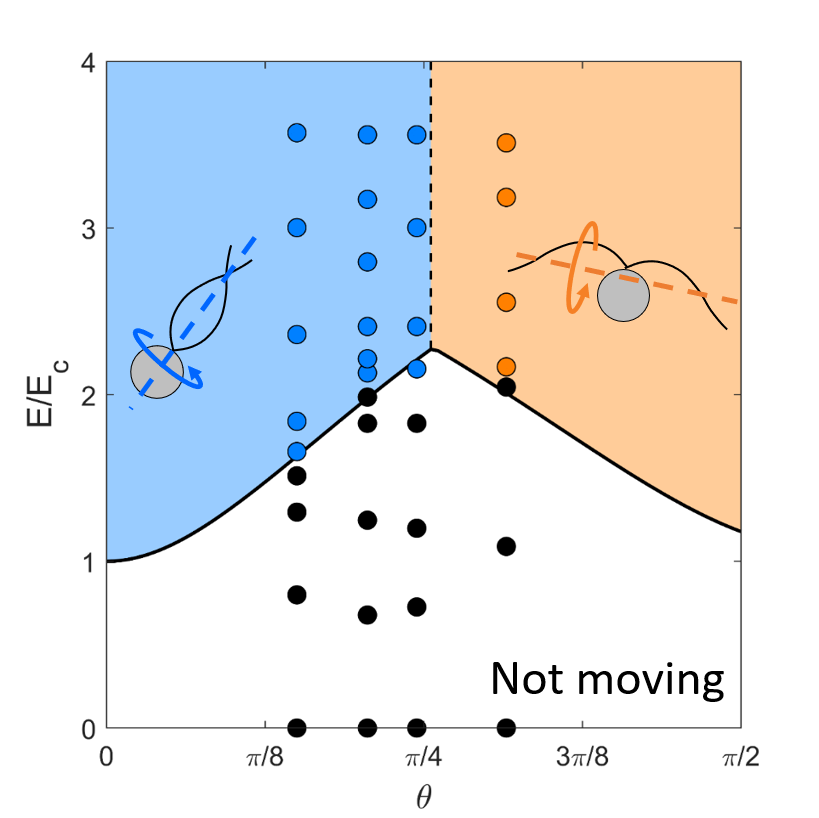}
    \end{center}
\caption{ State diagram for a smaller swimmer with sphere radius $R = 1.59$~mm, tail radius $a = 13.7~\mu$m, and tail length to sphere radius ratio $L/R = 6.4$. 
The blue points represent symmetric rotation, the orange points represent asymmetric rotation, and the black points are stationary. }
\label{SI_SmallSwimmer_StateDiagram}
\end{figure}

Besides the swimmers with sphere radius 3.18~mm and fiber radius 24.8~$\mu$m, we performed experiments with smaller swimmers as well with sphere radius $R = 1.59$~mm and fiber radius $a = 13.7$~$\mu$m. 
The materials of the sphere (HDPE) and the fiber (nylon) were the same as the bigger swimmers. 
The rotation of the small swimmers were qualitatively identical to the big ones. 
Fig.~\ref{SI_SmallSwimmer_AngularSpeed} shows the angular speed as a function of the applied field for a small swimmer with a tail angle $\theta = 43^\circ$ and length ratio $L/R = 5.8$.  
This swimmer rolled with its tails buckled toward the axis of symmetry.  
There was a significant hysteresis when the field decreased. 
However, for the small swimmers the oscillation state was either unstable or non-exist. 
The state diagram for the small swimmers with $L/R = 6.4$ and different tail angles $\theta$ is shown in Fig.~\ref{SI_SmallSwimmer_StateDiagram}. 
The boundaries were calculated using the same method introduced in Section~\ref{sec:Onset}, using $R = 1.59$~mm, $a = 13.7$~$\mu$m, and $\text{ln} \epsilon^{-1} = 6.6$.

\section{Torques generated by the tails}
\label{sec:Torque}

\begin{figure*}
    \begin{center}
        \includegraphics[scale=0.9]{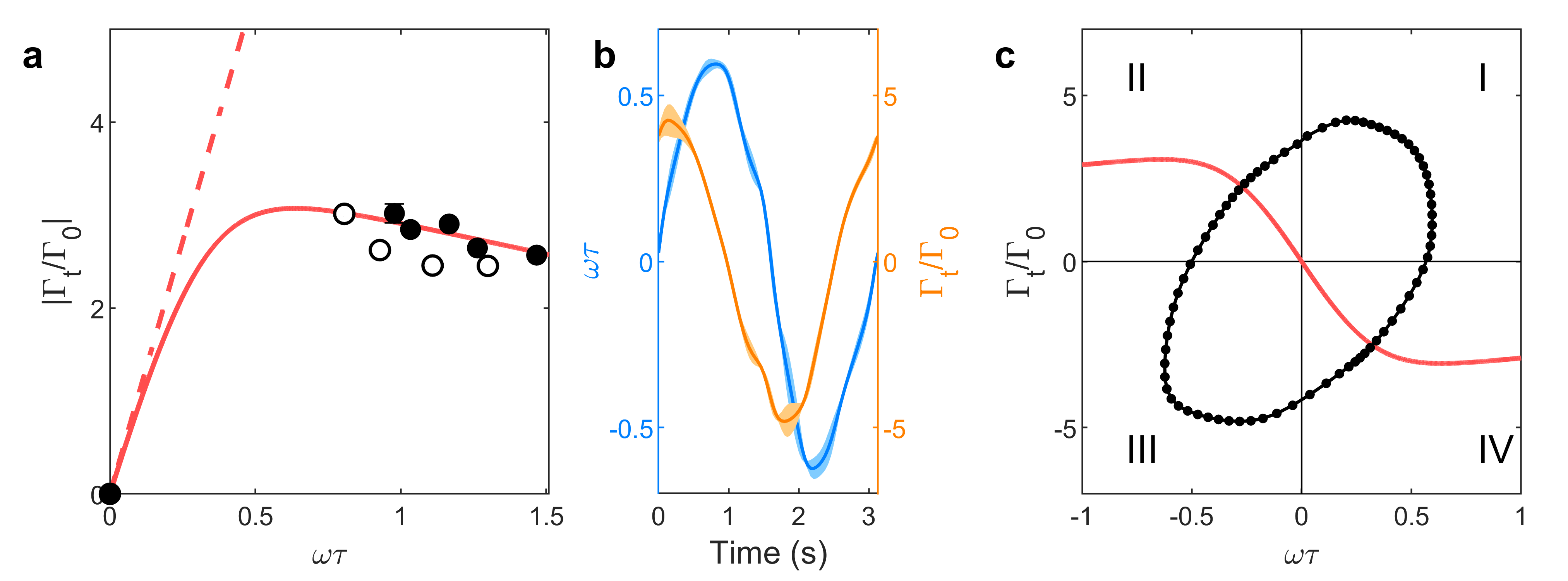}
    \end{center}
\caption{(a) Torque on one tail at different rotation speeds during steady rolling for the swimmer with $\theta = 47^\circ$, which is used as an example in Fig.~\ref{AngularSpeed} of the main text. Solid circles represent increasing $E$, and open circles represent decreasing $E$. The solid red curve is a theoretical estimation of $|\Gamma_\text{t}/\Gamma_0|$. The dashed red line shows its slope at $\omega \tau = 0$.   
(b) Angular speed (blue) and torque on the tails (orange) in one period of oscillation for the same swimmer at the oscillatory state. Shaded areas show standard deviations.  
(c) The trajectory of torque versus angular velocity when the sphere oscillated. The red curve is the steady-state response obtained in (a), but with a flipped sign. }
\label{SI_Torque}
\end{figure*}

\begin{figure*}
    \begin{center}
        \includegraphics[scale = 0.9]{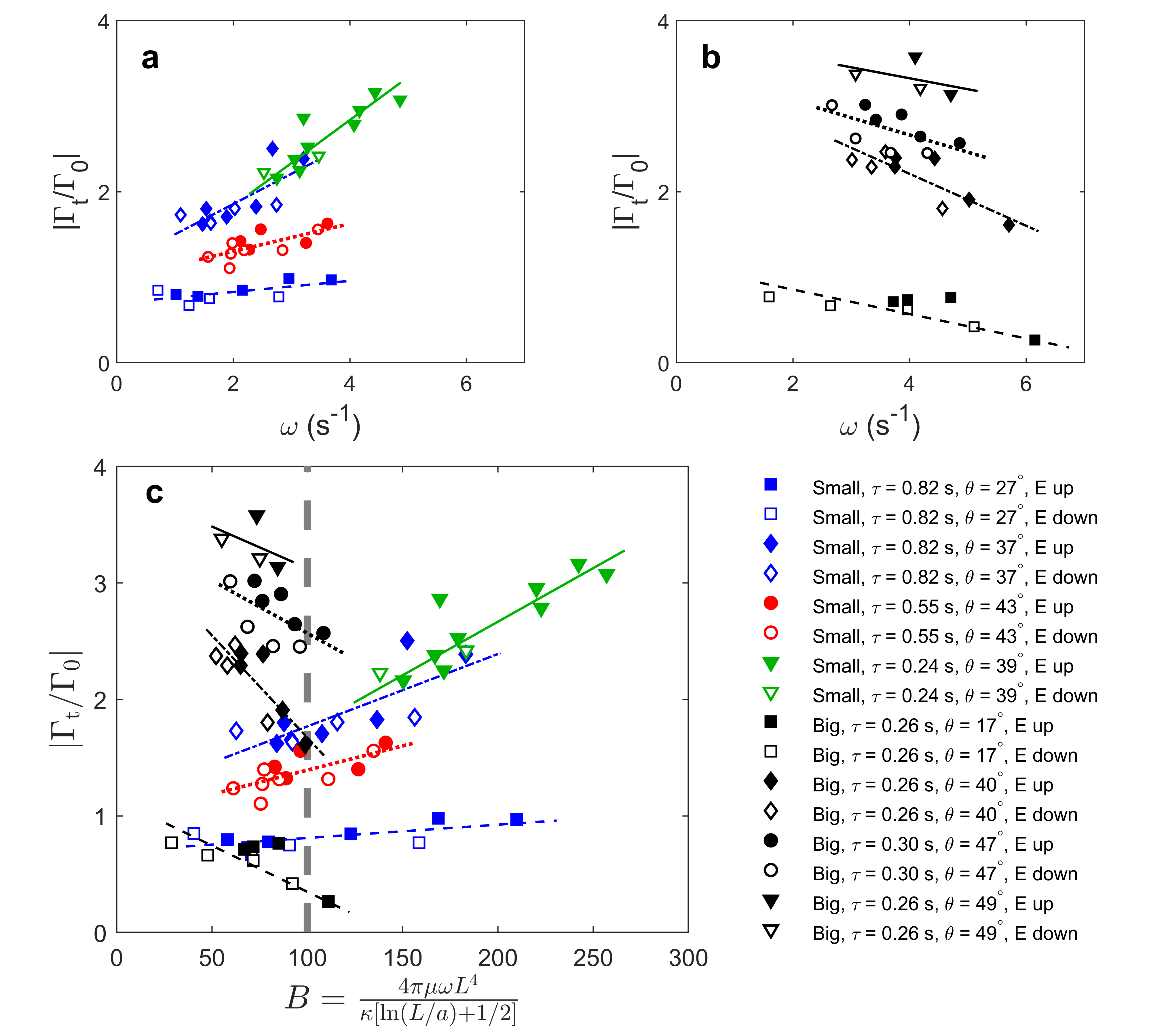}
    \end{center}
\caption{ Torques applied on the tails $\Gamma_\text{t}$ as functions of angular speed $\omega$ for different swimmers in liquids with different electrical relaxation time $\tau$ when the swimmers were rolling. 
Here $\Gamma_\text{t}$ is nondimensionalized by $\Gamma_0 = \kappa/L$. 
(a) Small swimmers with particle radius $R = 1.59$~mm and tail radius $a = 13.7~\mu$m. 
(b) Big swimmers with particle radius $R = 3.18$~mm and tail radius $a = 24.8~\mu$m. 
(c) Data in (a) and (b) plotted using the dimensionless bending number $B$ (Eq.~(\ref{eq:Sp}) in the main text) as the horizontal axis. The dashed grey line at $B = 10^2$ labels the approximate boundary between the data with positive and negative slopes. 
The symbols have the same meaning throughout (a-c). 
The straight lines are to guide the view. 
}
\label{SI_TorqueRescale}
\end{figure*}

In this section, we look at the torques generated by the swimmers' tails under different conditions. 
The total torque applied on the sphere always vanishes. 
As discussed in Section~\ref{sec:QuinckeTheory}, for a bare sphere the two torques involved are the driving torque due to the Quincke effect $\Gamma_\text{Q}$ and the resistance due to the particle's spin $\Gamma_\mu$, and 
\begin{equation}
    \Gamma_\text{Q}+\Gamma_\mu = 0. 
\end{equation}
For the bi-flagellated swimmers, we also need to consider the torque due to eccentric rotation of the sphere $\Gamma_\text{ec}$ and the torques generated by the two tails $\Gamma_\text{t1}$ and $\Gamma_\text{t2}$, so 
\begin{equation}
    \Gamma_\text{Q} + \Gamma_\mu + \Gamma_\text{ec} + \Gamma_\text{t1} + \Gamma_\text{t2} = 0. 
    \label{eq:SI_TorqueBalance}
\end{equation}
For the three stable forms of motion (rolling, pitching, and oscillation) discussed in the main text, the two tails are always symmetric, so $\Gamma_\text{t1} = \Gamma_\text{t2} = \Gamma_\text{t}$. 
Replace Eq.~(\ref{eq:DE_original}c) with Eq.~(\ref{eq:SI_TorqueBalance}), and substitute in Eq.~(\ref{eq:SI_T_nu}) and Eq.~(\ref{eq:SI_T_o}), we get 
\begin{subequations}
\begin{align}
    & \dot{p}_x = -\omega p_y - (p_x - p_0)/\tau, \label{eq:DE_swimmer_1} \\
    & \dot{p}_y = \omega p_x - p_y/\tau, \label{eq:DE_swimmer_2} \\
    & 0 = -E_0 p_y - 8 \pi \mu R^3 \omega - 6 \pi \mu R d^2 \omega + 2\Gamma_\text{t} \label{eq:DE_swimmer_3}
\end{align}
\label{eq:DE_swimmer}%
\end{subequations}
for our bi-flagellated swimmer, where $d$ is the distance between the center of the sphere and the axis of rotation. 
We assume the external field is in the $\vec{x}$ direction, and we assume no interaction between the electric field and the tails, so $\Gamma_\text{t}$ is contributed solely by the elastohydrodynamic interactions. 

Eqs.~(\ref{eq:DE_swimmer}) can be written with non-dimensional variables by defining $\mathcal{P}_x \equiv p_x/p_0$, $\mathcal{P}_y \equiv p_y/p_0$, $\tilde{t} \equiv t/\tau$, and $\tilde{\omega} \equiv \omega \tau$. At the same time, we define a characteristic torque $\Gamma_0 \equiv \kappa/L$ \cite{Fermigier_2008}, where $\kappa$ is the bending stiffness of the tail fiber, and $L$ is the tail length. 
Then Eqs.~(\ref{eq:DE_swimmer}) become 
\begin{subequations}
\begin{align}
    & \frac{d \mathcal{P}_x}{d\tilde{t}} = -\tilde{\omega} \mathcal{P}_y - \mathcal{P}_x + 1, \label{eq:DE_1}\\
    & \frac{d \mathcal{P}_y}{d\tilde{t}} = \tilde{\omega} \mathcal{P}_x - \mathcal{P}_y, \label{eq:DE_2} \\
    & 0 = \frac{8\pi \mu R^3 L}{\tau \kappa} \left[ \left( \frac{E}{E_c} \right)^2 \mathcal{P}_y - \tilde{\omega} -\frac{3}{4}\left( \frac{d}{R} \right)^2 \tilde{\omega} \right] + 2\frac{\Gamma_\text{t}}{\Gamma_0},  \label{eq:DE_3}
\end{align}
\label{eq:DE}%
\end{subequations}
where the prefactor $\dfrac{8\pi \mu R^3 L}{\tau \kappa}$ is similar to the elastoelectroviscous parameter $\bar{\mu}$ defined in \cite{Lailai_2019,Lailai_2020}. 
The parameters in the prefactor have all been measured experimentally. 

With Eqs.~(\ref{eq:DE}) and experimentally measured $E$, $d$, and $\omega(t)$, we can calculate $\Gamma_\text{t}(t)$.  
For rolling and oscillation, the axis of rotation roughly goes through the center of the sphere, so $d \approx 0$ and $\Gamma_\text{ec} \approx 0$. 
$\Gamma_\text{ec}$ needs to be considered in the case of pitching. 
As mentioned in Section~\ref{sec:QuinckeTheory}, $\Gamma_\text{t}$ is different when the swimmer rotates unidirectionally or oscillates. 
In the following discussion, we focus on rolling and oscillation, and calculate the corresponding $\Gamma_\text{t}$ separately.




When the sphere rolls with a constant angular speed, $\text{d}\mathcal{P}_x / \text{d}\tilde{t} = \text{d}\mathcal{P}_y / \text{d}\tilde{t} = 0$ and $d = 0$. 
Eqs.~(\ref{eq:DE}) lead to 
\begin{equation}
    \frac{\Gamma_\text{t}}{\Gamma_0} = -\frac{4\pi \mu R^3 L}{\kappa \tau} \left[ \left( \frac{E}{E_c} \right)^2 \frac{\tilde{\omega}}{1+\tilde{\omega}^2} - \tilde{\omega} \right]. 
    \label{eq:T_t}
\end{equation}
In this case, $\Gamma_\text{t}$ always resists rotation. 
Still, we use the swimmer with $\theta = 47^\circ$ as an example. 
The magnitudes of $|\Gamma_\text{t}/\Gamma_0|$ corresponding to the experimental measurements are plotted in Fig.~\ref{SI_Torque}(a). 
When the swimmer was rolling, $|\Gamma_\text{t}/\Gamma_0|$ was approximately a linear function of $\tilde{\omega}$ within $0.8 \lesssim \tilde{\omega} \lesssim 1.5$. 
In the limit of $\tilde{\omega} \to 0$, the asymptote labeled by the dashed red line represents the viscous torque applied on each tail assuming the fiber is straight and rigid when the swimmer rolls.  
The solid red curve is an approximation that smoothly connects these two linear regimes. 
As a comparison, in Fig.~\ref{AngularSpeed}(b) of the main text, we plotted $|\Gamma_\mu+2\Gamma_\text{t}|$, which has two linear regimes as well because $\Gamma_\mu$ is always a linear function of $\tilde{\omega}$. 
The experimental results are omitted in Fig.~\ref{AngularSpeed}(b), but that curve was calculated in a similar way. 

When the swimmer oscillates, $\Gamma_\text{t}$ is no longer controlled solely by the instantaneous angular speed $\omega$. 
We calculated $\Gamma_\text{t}$ numerically using experimentally measured $\omega(t)$ and Eqs.~(\ref{eq:DE}). 
We present the average $\tilde{\omega}$ and the corresponding $\Gamma_\text{t}/\Gamma_0$ in Fig.~\ref{SI_Torque}(b), and the trajectory of $\Gamma_\text{t}(\tilde{\omega})$ in Fig.~\ref{SI_Torque}(c), when the swimmer reached a steady oscillatory state. 
This oscillation can not be achieved if $\Gamma_\text{t}/\Gamma_0$ only follows the red curve shown in Fig.~\ref{SI_Torque}(a). 
However, at larger $E$, the trajectory will eventually migrate to and reside on the red curve and the swimmer rotates unidirectionally.

Lastly, we look at the torques applied on the tails for swimmers with different geometries (sphere radius $R$, tail radius $a$, tail length $L$ and tail angles $\theta$) in liquids with different electrical properties (mainly the relaxation time $\tau$). 
When the swimmers are rolling, the motion of the tails is identical to those described in Refs.~\cite{Manghi_PRL,Breuer_2008,Fermigier_2008,Stone_Chapter_2012}. 
As shown in Fig.~\ref{SI_TorqueRescale}, $|\Gamma_\text{t}/\Gamma_0|$ increased with $\tilde{\omega}$ for the small swimmers (Fig.~\ref{SI_TorqueRescale}(a)), and decreased for the big swimmers (Fig.~\ref{SI_TorqueRescale}(b)). 
Following the spirit of~\cite{Breuer_2008,Fermigier_2008}, we plot the rescaled data using the dimensionless bending number 
\begin{equation}
    B = \frac{4\pi \mu \omega L^4 }{\kappa [\mathrm{ln}(L/a)+1/2]} 
\end{equation}
as the horizontal axis. 
There is a transition at $B \approx 10^2$. 
When $B < 10^2$, $|\Gamma_\text{t} / \Gamma_0|$ decreased with $B$, and when $B > 10^2$, $|\Gamma_\text{t} / \Gamma_0|$ increased with $B$. 
This agrees qualitatively with the conclusions in Refs.~\cite{Breuer_2008, Fermigier_2008}, where an s-shaped curve of $B$ versus $|\Gamma_\text{t}/\Gamma_0|$ is predicted for rotating fibers with a fixed angle $\theta$. 
Based on their results, the slope of $|\Gamma_\text{t}|$ is positive when $B$ is below a second threshold on the left-hand side of $B = 100$. 
As $B$ increases from 0, the torque increases first, then decreases, and lastly increases again. 
What our experiments explored are the second and third regimes, and the transition happened at about $B \approx 100$ according to the data.

\end{document}